\documentclass[a4paper,fleqn,usenatbib]{mnras}

\usepackage{newtxtext,newtxmath}

\usepackage[T1]{fontenc}
\usepackage{ae,aecompl}

\usepackage{graphicx}	
\usepackage{amsmath}	
\usepackage{amssymb}	

\title[SRT observations of LG dwarf galaxies]{Sardinia Radio Telescope observations of Local Group dwarf galaxies -- I. The cases of NGC\,6822, IC\,1613, and WLM}
\author[A. Tarchi et al.]{
A. Tarchi,$^{1}$\thanks{E-mail: andrea.tarchi@inaf.it}
P. Castangia,$^{1}$
G. Surcis,$^{1}$
A. Brunthaler,$^{2}$
C. Henkel,$^{2,3}$
\newauthor 
M. Pawlowski,$^{4}$
K. M. Menten,$^{2}$
A. Melis,$^{1}$
S. Casu,$^{1}$
M. Murgia,$^{1}$
\newauthor 
A. Trois,$^{1}$
R. Concu,$^{1}$
and J. Darling$^{5}$
\\
$^{1}$INAF-Osservatorio Astronomico di Cagliari, Via della Scienza 5, 09047 Selargius (CA), Italy\\
$^{2}$Max-Planck-Institut f{\"u}r Radioastronomie, Auf dem H{\"u}gel 71,53121 Bonn, Germany\\
$^{3}$Astronomy Departement, King Abdulaziz University, P.O. Box 80203, Jeddah 21589, Saudi Arabia\\
$^{4}$Leibniz-Institut f\"{u}r Astrophysik Potsdam, An der Sternwarte 16, 14482 Potsdam, Germany\\
$^{5}$Center for Astrophysics and Space Astronomy, Department of Astrophysical and Planetary Sciences,\\
University of Colorado, 389 UCB, Boulder, CO 80309-0389, USA
}

\date{Accepted XXX. Received YYY; in original form ZZZ}

\pubyear{2015}

\begin{document}

\newcommand{\HI}{H\kern0.1em{\sc i}}
\newcommand{\hii}{H\kern0.1em{\sc ii}}
\newcommand{\txs}{TXS\,2226{\tt -}184}
\newcommand{\pks}{PKS\,2322{\tt -}123}
\newcommand{\radm}{rad m$^{-2}$}
\newcommand{\ab}{$\sim$}
\newcommand{\etal}{{\sl et~al.\ }}
\newcommand{\dg}{$^{\circ}$}
\newcommand{\kms}{km s$^{-1}$}
\newcommand{\solmass}{\hbox{M$_{\odot}$}}
\newcommand{\solum}{\hbox{L$_{\odot}$}}

\label{firstpage}
\pagerange{\pageref{firstpage}--\pageref{lastpage}}
\maketitle

\begin{abstract}
Almost all dwarf galaxies in the Local Group that are not satellites of the Milky Way or M\,31, belong to either one of two highly-symmetric planes. It is still a matter of debate, whether these planar structures are dynamically stable or whether they only represent a transient alignment. Proper motions, if they could be measured, could help to discriminate between these scenarios. Such motions could be determined with multi-epoch Very Long Baseline Interferometry (VLBI) of sources that show emission from water and methanol at frequencies of 22 and 6.7 GHz, respectively. We report searches for such masers. We have mapped three Local Group galaxies, NGC 6822, IC 1613 and WLM in the bands covering the water vapor and methanol lines. These systems are members of the two above mentioned planes of galaxies. We have produced deep radio continuum (RC) maps and spectral line cubes. The former have been used to identify star forming regions and to derive global galactic star formation rates (SFRs). These SFRs turn out to be lower than those determined at other wavelengths in two of our sources. This indicates that dwarf galaxies may follow predictions on the RC-SFR relation only in individual regions of enhanced radio continuum emission, but not when considering the entire optical body of the sources. No methanol or water maser emission has been confidently detected, down to line luminosity limits of $\sim$ 4$\times10^{-3}$ and  10$\times10^{-3}$ \solum, respectively. This finding is consistent with the small sizes, low SFRs and metallicities of these galaxies.
\end{abstract}




\section{Introduction}\label{sect:intro}
The dwarf galaxies in the Local Group (LG), the only galaxy ensemble for which accurate 3D positions are known, reveal a surprising amount of spatial structure. The satellite galaxies of the Milky Way (MW) lie within a vast polar structure (VPOS), a thin (28 kpc root-mean-square - rms - height) plane extending to a Galactocentric distance of 250 kpc (\citealt{pawlowski2012}). Similar planes of satellites surround M31 (\citealt{ibata2013}) and Centaurus A (\citealt{muller2018}). In addition to these satellite galaxy planes, the distribution of dwarf galaxies in the LG reveals further structuring (\citealt{pawlowski2013a}). Almost all other LG dwarfs belong to either one of two planes, termed LG plane 1 (10 galaxies) and LG plane 2 (7 galaxies). The most puzzling feature of the last two planes is their pronounced symmetry. Both planes have large diameters (1.5 to 2 Mpc), but are extremely thin (32.5 $\pm$ 1.9 and 59.9 $\pm$ 4.0 kpc, respectively). They both are inclined by only $\approx 19$ degrees with respect to the stellar disc of M31, but by 30 degrees relative to each other. In addition, both planes (1 and 2) have essentially the same offset ($\approx 190$~kpc) from the MW and from M31, i.e. they are both almost parallel to the line connecting the MW and M31.
If all these structures in the LG were dynamically stable then the galaxies would move within the associated plane, which would allow us to predict the expected proper motion of individual satellites in such structures (\citealt{pawlowski2013b}, \citealt{pawlowski2015}). Otherwise, the structures may only represent transient alignments that would disperse as the dwarf galaxies continue their independent motions, as expected from cosmological simulations (e.g., \citealt{gillet2015}, \citealt{buck2016}). For the VPOS of the MW, this can be tested using optical proper motion measurements, revealing that indeed most of the classical MW satellites co-orbit in the (VPOS) plane (\citealt{metz2008}, \citealt{pawlowski2013b}, \citealt{helmi2018}, \citealt{fritz2018}). Likewise, the favorable edge-on orientation of the M31 satellites as seen from the MW allowed (\citealt{ibata2013}) to use only radial velocities (hereafter, to be understood as line-of-sight velocities) to demonstrate that 13 of the 15 satellites in the plane are consistent with co-orbiting motion. Unfortunately, information on the motions of dwarf galaxies belonging to the two LG planes (1 and 2) is not easily available. These planes are not seen edge-on, rendering radial velocities alone inconclusive, and optical proper motion measurements are currently not feasible due to the large distances of these objects.

For nearby galaxies, a viable method to derive proper motions is offered by VLBI studies of the 22-GHz water maser line in star-forming regions. Indeed, VLBI observations of water maser spots in phase-referencing mode allow us to measure the 3D motions of the host galaxies with respect to distant quasars. In addition, distance measurements are possible by applying the rotational parallax method to the detected maser spots with relevant implications for cosmological parameters and the total mass of matter (luminous and dark) of the LG. So far, such studies have been successfully performed on a limited number of maser spots detected in the galaxies M\,33 and IC\,10 (e.g. \citealt{greenhill1993}; \citealt{brunthaler2005}; \citealt{brunthaler2007}). In particular, the measurement performed on IC 10 shows the potential lying in such studies for our case. This galaxy belongs to the LG plane 2, with a distance from the best-fit plane of only 37 $\pm$ 13 kpc (0.62 $\pm$ 0.22 rms heights). In addition to its radial velocity, IC 10's proper motion has been measured using water masers (\citealt{brunthaler2007}). The resulting velocity vector of the galaxy aligns to better than 10 degrees with the LG plane 2, demonstrating the possible dynamical association of this dwarf galaxy with its plane. Alternatively, also methanol maser lines can be used for proper motion studies. This possibility, for extragalactic studies, has been further supported by the detection of methanol maser sources in M31 (\citealt{sjouwerman2010}).\\

Within the aforementioned framework, it is particularly important to detect as many water/methanol maser sources as possible in any galaxy belonging to the LG and, in particular, in the LG dwarf galaxies belonging to the LG planar structures 1 and 2. Until recently, water maser sources in the LG were found only in M\,33 (3 sources), IC\,10 (2 sources), the LMC (30 sources) and the SMC (6 sources; see \citealt{henkel2018}, and references therein). Possible causes for this low-detection rate can be both the difficulty to map the entire body of the galaxies and the limited sensitivity of the observational campaigns. The former motivation is especially dramatic for M\,31 that covers (due to its size and proximity) a huge area of the sky (120$^{\prime} \times 40^{\prime}$). Indeed, at the end of 2010, the first detection of maser emission in five (out of 206) 24 $\mu$m selected regions in M31 has been obtained with the Green Bank Telescope, GBT (\citealt{darling2011}). A few months earlier, the detection of methanol maser emission was reported by Sjouwerman et al. (2010) in the disk of M31. However, the lack of a counterpart at another wavelength was sometimes reported in maser detections (see, e.g., \citealt{becker1993} for IC\,10-NW; \citealt{amiri2016} for one case in M\,31). This strongly motivates un-biased searches for maser sources in fully-sampled spectroscopic images rather than focusing only on well-known isolated hotspots.

This paper reports the first results of a large project, carried out with the Sardinia Radio Telescope (SRT), aimed at mapping the radio continuum and spectral line emission at C ($\lambda \sim$ 5 cm) and K ($\lambda \sim$ 1.3 cm) band from 14 (out of a total of 17) Local Group dwarf galaxies belonging to the LG planar structures 1 and 2, and being visible at the SRT latitude (and excluding IC\,10). The final aim of the project is that of obtaining spectro-polarimetric maps of the full optical body\footnote{Defined by the major angular diameter in arcminutes, corresponding to the Holmberg isophote (26$^m$.5\,arcsec$^{-2}$) in the B band (from \citealt{karachentsev2013}).} of the galaxies with the SRT at a resolution of about 3$^\prime$ and 1$^\prime$, at C and K band, respectively. The main scientific goals are: i) to detect spectral line emission in the observed band, in particular from the 6.7-GHz methanol and 22-GHz water maser lines and to locate them with high-enough precision for interferometric follow-up studies; ii) to obtain deep and full polarization maps of the radio continuum emission. These will provide relevant clues on individual star forming regions hosting groups of thermal (compact or regular HII regions) and non-thermal (radio supernovae and supernova remnants) sources present in the target galaxies. The present work reports the details of the observations and data reduction (Sect.~\ref{sect:obs}), and summarizes the main results achieved on the first three galaxies analyzed, so far, NGC\,6822, IC\,1613, and WLM (Sect.~\ref{sect:results}). Sect.~4 provides a discussion on the interpretation of the intensity and distribution of the radio continuum sources, and of the number of maser detections found in the three targets. The last section itemizes the main conclusions produced by the first part of this large project. The results for the other galaxies in the sample, together with a more comprehensive study of the sample, will be the subject of a subsequent paper (Paper II).

\section{Observations and Data Reduction}\label{sect:obs}

Observations were performed with the SRT (\citealt{bolli2015}; \citealt{prandoni2017}, and references therein) in the framework of the Early Science Program (ESP) which started on February 1, 2016 and lasted for about six months.

Our project, labeled ESP-S0003 (220 hours; PI: A. Tarchi), aimed at mapping the full extension of 14 (out of a total of 17) Local Group dwarf galaxies belonging to the LG planar structures 1 and 2 that were visible at the SRT latitude (excluding IC\,10). We used the C and K band receivers\footnote{At K band, only the central feed of the 7-feed receiver was used in our observations, since, at that time, only this feed was offered in conjunction with the SARDARA backend.} in conjunction with the SARDARA, ROACH2-based digital backend (\citealt{melis2018}). At C-band we observed with an actual bandwith (BW) of 1.2 GHz centered around 6.6 GHz, in order to cover both methanol and excited OH maser line transitions (near 6.7 and 6.0 GHz, respectively), while trying to avoid strong RFI present at 5.9-GHz. At K-band, the 1.2-GHz actual band was centered close to (but not coincident with\footnote{We centered the band with an offset of 15 MHz toward lower frequencies, with respect to (w.r.t.) the frequency of the main water maser line Doppler-shifted by the systemic heliocentric velocity of the targets. This was done in order to avoid to center the possible systemic line on a strong birdie, known to be present in the central channel.}) the frequency of the main water maser line (rest frequency 22.23508 GHz) Doppler-shifted by the systemic heliocentric velocity of the targets (taken by \citealt{mcconnachie2012}). We used the 1.5-GHz (nominal) band SARDARA configuration with 16384 channels, yielding a frequency resolution of $\sim$ 91 kHz ($\approx$ 4 km/s and 1.2 km/s, at C and K band, respectively).

The spectroscopic check of the system was done by observing, in position-switching mode, the strong methanol and water maser emission in the well-known source W3(OH).
 
All maps were performed using the On-The-Fly (OTF) mapping technique using scan speeds between 3\dg/min (at C band) and 1-3 \dg/min (at K band, depending on the size of the map; the smaller the map, the lower the speed). 
At C band we mapped, for all galaxies, square regions of 21$^\prime$ $\times$ 21$^\prime$ centered on the target coordinates (J2000 epoch), as reported in the NASA/IPAC Extragalactic Database (NED). The area was covered with 29 scans (in each direction) separated by 45 arcsec along both RA and DEC, with an acquisition rate of $\sim$ 30 spectra/sec, thus properly sampling the $\sim 3^\prime$ SRT beam at C band. At K band the size of the regions observed depends on the optical size of the galaxy. For NGC\,6822 and IC\,1613, we mapped a region of 18$^\prime$ $\times$ 18$^\prime$ with a 15 arcsec separation along both RA and DEC (thus 73 scans, in each direction), while for WLM a 12$^\prime$ $\times$ 12$^\prime$ sized map with a 15 arcsec separation (49 scans, in each direction) was performed. The size of the maps at C band included the full optical body of all three galaxies (see Table\ref{table:targets}). At K-band, the whole optical extent of the galaxy WLM was mapped, while for the most extended targets, NGC\,6822 and IC\,1613, the maps were slightly smaller with respect to (w.r.t.) the optical major axis, viz. 18$^{\prime}$ vs.\ 19$^{\prime}$ (see Table\ref{table:targets}). The acquisition rate was $\sim$ 60 spectra/sec sufficient to have a  proper sampling of the  $\sim 1^\prime$ SRT beam at K band. Multiple maps along RA and DEC were performed for each galaxy that were later averaged to increase the final sensitivity. 
Table~\ref{table:obs} reports all the details of the observations.

The data were calibrated by using the proprietary Single-dish Spectral-polarimetry Software (SCUBE; \citealt{murgia2016}). The name of the observed calibrators are reported in Table~\ref{table:obs}, together with an indication on which of them were used for the flux and/or polarization calibration. The steps of the data reduction, including flux density, polarization calibration, and RFI flagging for a standard spectro-polarimetric on-the-fly SRT data set are thoroughly described in Murgia et al. (2016; their Sect. 3) and Govoni et al. (2017; their Sect. 3.1). Hence, with the exception of possible peculiar problems related to the data reduction of our project, the steps will not be repeated here. In addition, with SCUBE we have also performed total intensity and polarization imaging using the methods described in \citet[their Sects. 4.1 and 4.3]{murgia2016} and \citet[their Sects. 3.1.1 and 3.1.2]{govoni2017}. Unfortunately, the polarization calibration at K band failed, due to problems under investigation, and hence, only polarization imaging at C band is reported and discussed in this paper. 

We created total intensity image cubes of all 16384 channels both at C and K band, and produced radio continuum images by averaging the inner 80\% of the whole bands, after removing the channels where birdies are known to be present. Cubes at different epochs were re-aligned in velocity using SCUBE, in order to account for the Doppler shift introduced by the Earth rotation. The proper velocities were computed by using 'offline' the FTrack software (Orlati et al. in prep.) developed for the Italian radio telescopes and not yet implemented for SARDARA backend acquisitions at the time of the observations. For each source, cubes at different epochs were finally averaged and the averaged cubes searched for methanol and water maser line features using the following method and criteria:
\begin{itemize}
\item we used the first step of the identification process described in \citet[their Sect. 3.1]{surcis2011}. This implies the use of a program called ``maser finder'' (developed by S. Curiel), which is able to search for maser spots, velocity channel by velocity channel, with a signal-to-noise ratio greater than a given value (in our case, a 5-sigma limit was used);
\item we limited our search to a frequency window corresponding to a range in velocity of $\pm$ 200 \kms, centered on the frequency at which the 6.7-GHz methanol and 22-GHz water maser line is expected due to the systemic heliocentric velocity of the galaxy (taken from \citealt{mcconnachie2012});
\item we collected all candidate maser features found in the process; 
\item we considered as detections those candidate maser features whose peak flux density exceeds by seven times the rms found in the spectrum at the position of the feature;
\item we considered as tentative detections those candidate maser features whose peak flux density was between four and seven times the rms found in the spectrum at the position of the feature (see also Appendix A).
\end{itemize}

\begin{table*}
\caption{Target name and coordinates, LG plane to which the galaxy seems to belong (see Sect.~\ref{sect:intro}), distance, and optical size.}
\label{table:targets}
\begin{tabular}{lccccc}
       &  &          &   &   \\
\hline
Source          & RA     & Dec    & LG plane &  D$^{a)}$      &   Optical size$^{b)}$  \\
                & (J2000)& (J2000)&          &  (kpc)  &     $^{\prime}$\\
\hline
NGC\,6822       & 19h44m57.7s & -14d48m12s & 2 & 490$\pm$40    &   19.05  \\ 
 IC\,1613       & 01h04m47.8s & +02d07m04s & 1 &  700$\pm$35  &   19.05  \\
 WLM            & 00h01m58.1s & -15d27m39s & 1 &  925$\pm$40  &   11.48   \\
\hline
\end{tabular}
%
\\
$^{a)}$ from Mateo (1998)\\
$^{b)}$ Major angular diameter in arcminutes, corresponding the Holmberg isophote in the B band (from \citealt{karachentsev2013}).
\end{table*}

\begin{table*}
\caption{Details of the SRT observations}             
\label{table:obs}      
\begin{tabular}{lccccl}        
\hline\hline     
Source    &      Frequency   &   Obs. Date  &        OTF Maps         & Mapping time$^{a)}$  &   Observed calibrators$^{b)}$\\
          &        (MHz)     &   (yyyy-mm-dd) &      ($n$RA$\times$$n$DEC)                 &  (hours)      &              \\ 
\hline  
NGC\,6822   &       6001-7201  &   2016-04-01 &     19$\times$19 &     5.0        &       3C\,48, {\it 3C\,286}, 3C\,295, NGC\,7027 \\

NGC\,6822   &       21574-22774  &   2016-03-30 &     8$\times$8 &     5.1        &       3C\,48, {\it 3C\,286}, 3C\,295, NGC\,7027 \\

NGC\,6822   &       21574-22774  &   2016-04-03 &     7$\times$7 &     4.5         &      3C\,48, {\it 3C\,286}, NGC\,7027 \\

NGC\,6822   &       21574-22774  &   2016-04-10 &     5$\times$5 &     3.2         &      3C\,48, 3C\,147, {\it 3C\,286}, NGC\,7027 \\

NGC\,6822   &       21574-22774  &   2016-04-13 &     9$\times$9 &     5.8         &       3C\,48, {\it 3C\,286}, 3C\,295, NGC\,7027 \\

NGC\,6822   &       21574-22774  &   2016-04-15 &     8$\times$8 &     5.1        &       3C\,48, {\it 3C\,286}, 3C\,295, NGC\,7027 \\

NGC\,6822   &       21574-22774  &   2016-04-16 &     9$\times$9 &     5.9         &       3C\,48, {\it 3C\,286}, 3C\,295, NGC\,7027 \\

NGC\,6822   &       21574-22774  &   2016-04-17 &     9$\times$9 &     5.8         &       3C\,48, {\it 3C\,286}, 3C\,295, NGC\,7027 \\

NGC\,6822   &       21574-22774  &   2016-04-19 &     5$\times$5 &     3.2        &       {\it 3C\,286}, NGC\,7027 \\
\hline
IC\,1613    &       6005-7202   &   2016-03-12 &    16$\times$16 &     4.2        &    3C\,138, {\it 3C\,147}, NGC\,7027 \\

IC\,1613    &       6005-7202   &   2016-04-01 &     8$\times$8 &      5.1        &    {\it 3C\,48}, 3C\,138, NGC\,7027 \\

IC\,1613    &       6042-7202   &   2016-04-02 &     5$\times$5 &      3.2         &    {\it 3C\,48}, 3C\,138, NGC\,7027 \\

IC\,1613    &       21588-22788 &   2016-04-13 &     6$\times$6 &      3.8         &    {\it 3C\,48}, NGC\,7027 \\

IC\,1613    &       21588-22788 &   2016-04-15 &     5$\times$5 &      3.2         &    {\it 3C\,48}, 3C\,138, NGC\,7027 \\

IC\,1613    &       21588-22788 &   2016-04-16 &     6$\times$6 &      3.8         &    {\it 3C\,48}, NGC\,7027 \\

IC\,1613    &       21588-22788 &   2016-04-17 &     6$\times$6 &      3.8        &    {\it 3C\,48}, NGC\,7027 \\

IC\,1613    &       21588-22788 &   2016-04-19 &     6$\times$6 &      3.8         &    {\it 3C\,48}, 3C\,147, NGC\,7027 \\

IC\,1613    &       21588-22788 &   2016-04-20 &     5$\times$5 &      3.2         &    {\it 3C\,48}, 3C\,147 \\

IC\,1613    &       21588-22788 &   2016-04-22 &     8$\times$8 &      5.1         &    {\it 3C\,48} \\
\hline
WLM    &       6002-7202   &   2016-03-13 &    5$\times$5 &    1.3        &  {\it 3C\,48}, 3C\,138, NGC\,7027 \\

WLM    &       6002-7202   &   2016-04-10 &    12$\times$12 &   3.2         &  3C\,84, 3C\,138, {\it 3C\,147} \\

WLM    &       21569-22769 &   2016-04-27 &     14$\times$14 &    6.0         &  3C\,138, {\it 3C\,147}, 3C\,295 \\

WLM    &       21569-22769 &   2016-05-24 &     9$\times$9 &    3.8         &  3C\,138, {\it 3C\,147}, NGC\,7027 \\

WLM    &       21569-22769 &   2016-05-25 &     11$\times$11 &     4.7          &   3C\,138, {\it 3C\,147}, NGC\,7027 \\
\hline
\end{tabular}\\
$^{a)}$ Including overheads (acceleration and deceleration during scans).\\
$^{b)}$ Sources used for flux calibration are marked in italics. Polariziation angle was calibrated using 3C\,286 (for NGC\,6822) and 3C\,138 (for IC\, 1613 and WLM). For all three targets, polariziation leakage was calibrated using NGC\,7027, with the exception of one K-band epoch of WLM, for which 3C\,84 was employed. 
\end{table*}

The analysis of the radio continuum and spectral line maps was mainly led utilizing standard tasks implemented in the NRAO Astronomical Image Processing System (AIPS\footnote{http://www.aips.nrao.edu}).

The overlays presented in Sect.~\ref{sect:continuum} were produced using the 'kvis' image/movie viewer in the KARMA software\footnote{https://www.atnf.csiro.au/computing/software/karma/index.html}. 

\section{Results}\label{sect:results}

\begin{figure*}
   \centering
   \includegraphics[width=8.5cm]{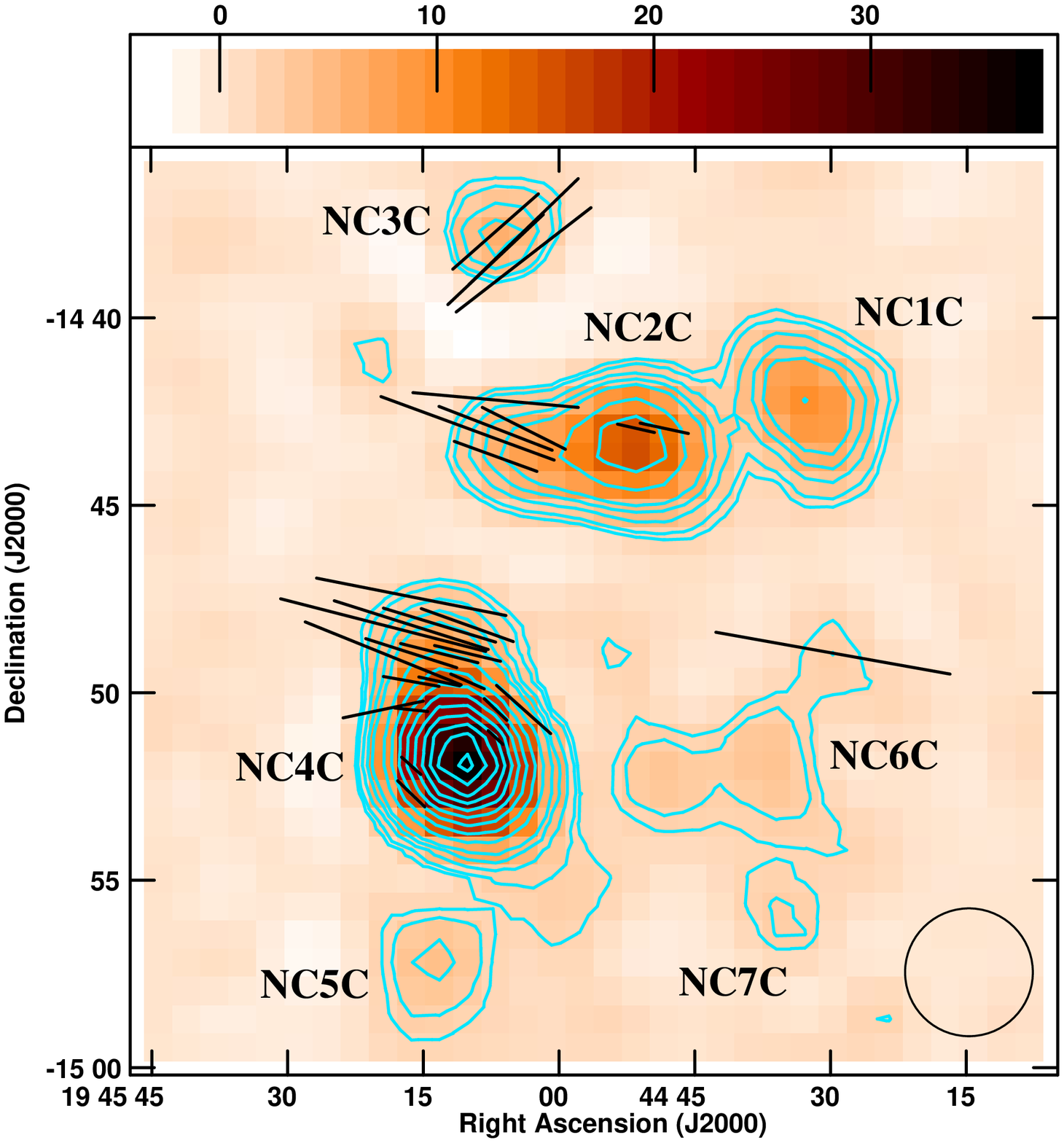}
   \includegraphics[width=8.5cm]{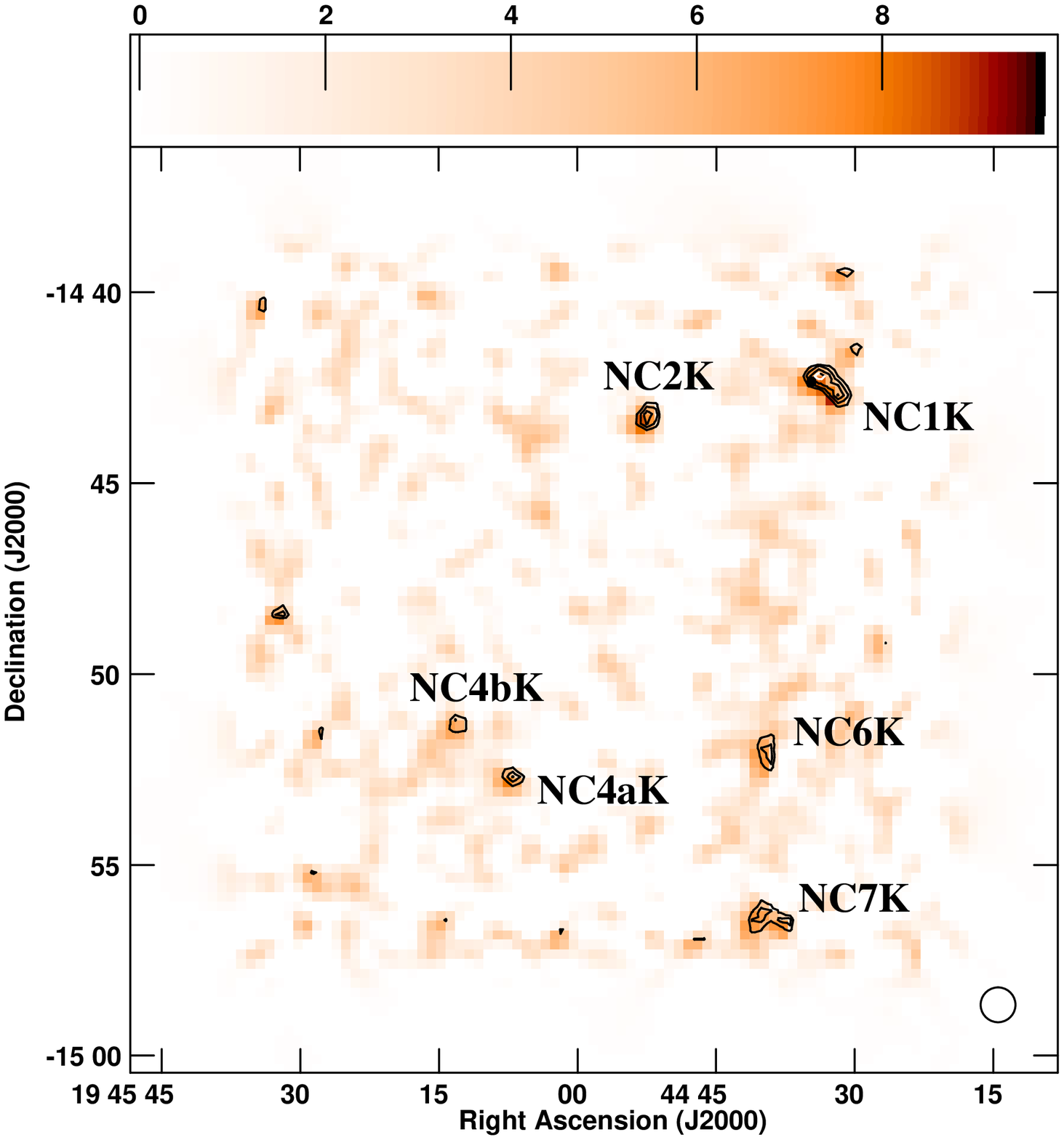}
      \caption{
{\it Left:} SRT C-band total intensity $21^{\prime}\times21^{\prime}$ image of NGC\,6822 resulting from the spectral average of the inner 80\% of the bandwidth. The FWHM beam is 2.9 arcmin (bottom-right corner). The noise level is 0.65 mJy/beam. Contour levels are 0.65 $\times$ (3, 4, 6, 8, 10, 15, 20, ..., 80) mJy. Electric field polarization vectors are overlaid. The length of the vectors is proportional to the polarization percentage (with 10\% being a bar of $\sim$ 2.1 arcmin), while their orientation represents the polarization angle. The error on the polarization angle is less than 10\dg, and the fractional polarization is above 3$\sigma_{FPOL}$.  {\it Right:} SRT K-band total intensity $18^{\prime}\times18^{\prime}$image of NGC\,6822 resulting from the spectral average of the inner 80\% of the bandwidth. The FWHM beam is 0.9 arcmin (bottom-right corner). The noise level is 2.0 mJy/beam. Contour levels are 2.0 $\times$ (3, 3.5, 4, 4.5, 5, 6, 7) mJy.
}
         \label{fig:n6822ck}
   \end{figure*}

\begin{figure*}
   \centering
   \includegraphics[width=8.5cm]{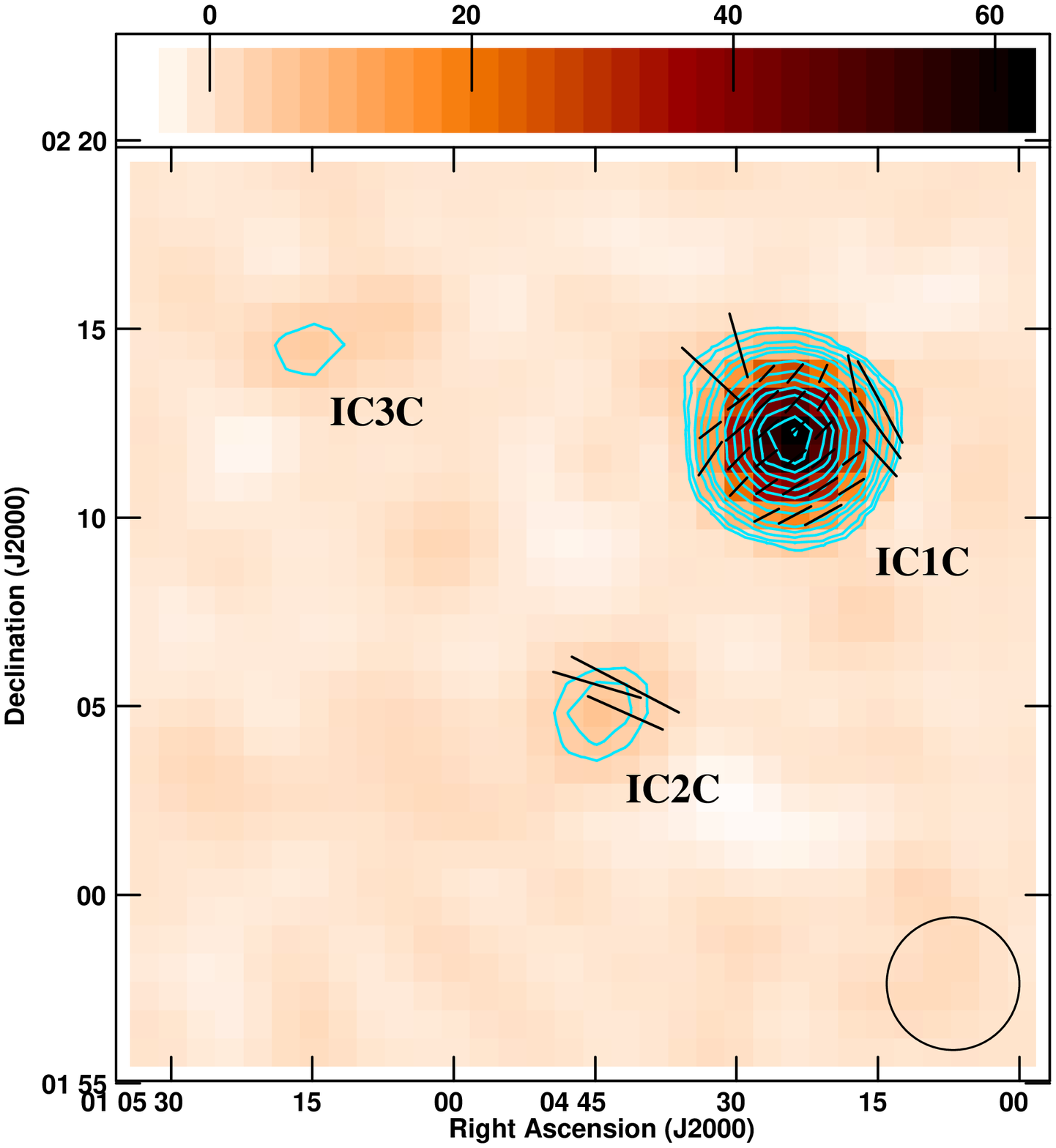}
   \includegraphics[width=8.5cm]{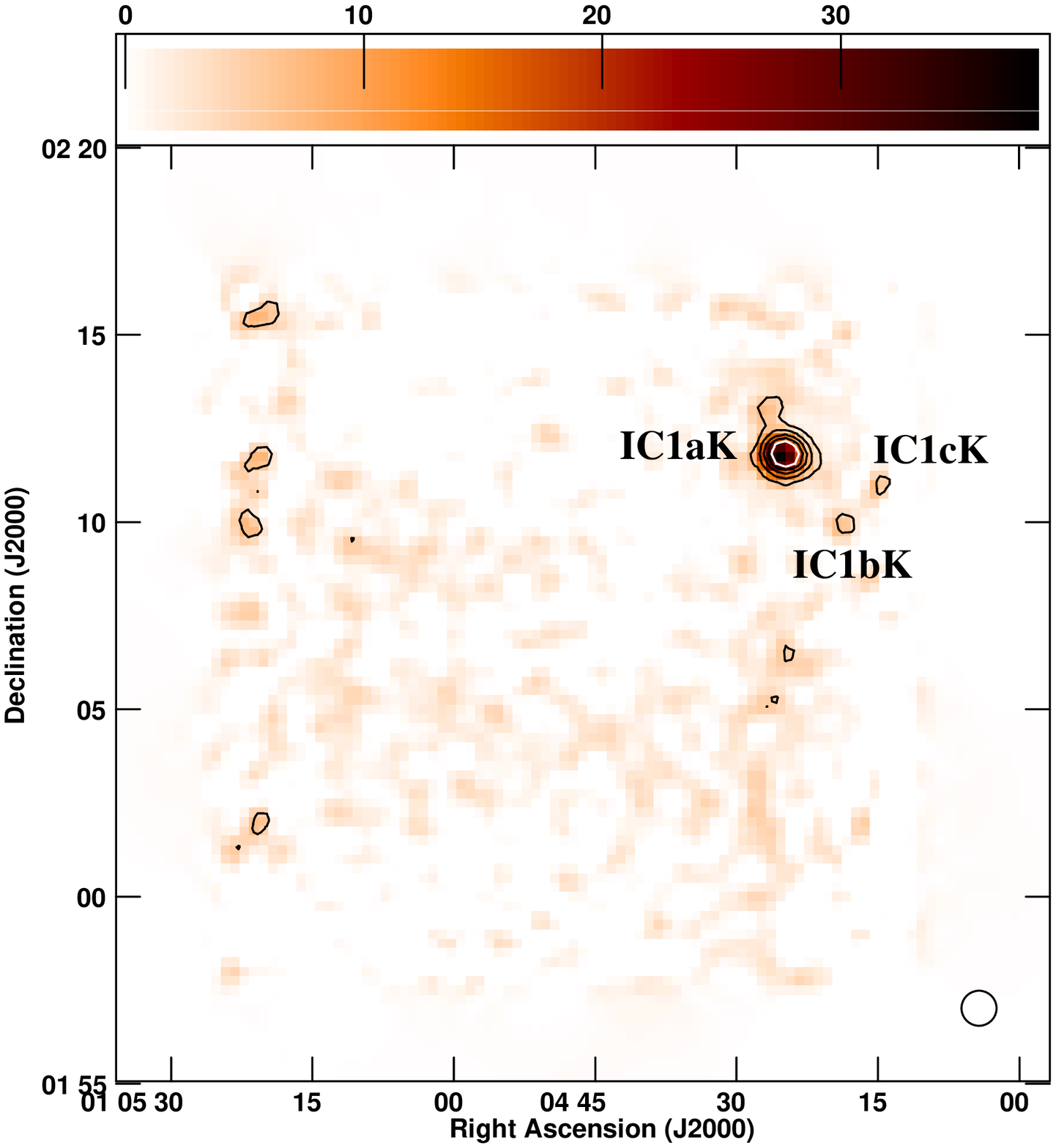}
      \caption{
{\it Left:} SRT C-band total intensity $21^{\prime}\times21^{\prime}$ image of IC\,1613 resulting from the spectral average of the inner 80\% of the bandwidth. The FWHM beam is 2.9 arcmin (bottom-right corner). The noise level is 1.0 mJy/beam. Contour levels are 1.0 $\times$ (3, 4, 6, 8, 10, 15, 20, ..., 80) mJy. Electric field polarization vectors are overlaid. The length of the vectors is proportional to the polarization percentage (with 10\% being a bar of $\sim$ 2.1 arcmin), while their orientation represents the polarization angle. The error on the polarization angle is less than 10\dg, and the fractional polarization is above 3$\sigma_{FPOL}$.  {\it Right:} SRT K-band total intensity $18^{\prime}\times18^{\prime}$ image of IC\,1613 resulting from the spectral average of the inner 80\% of the bandwidth. The FWHM beam is 0.9 arcmin (bottom-right corner). The noise level is 1.7 mJy/beam. Contour levels are 1.7 $\times$ (3, 6, 9, 12, 15) mJy.
              }
         \label{fig:ic1613ck}
   \end{figure*}

\begin{figure*}
   \centering
   \includegraphics[width=8.5cm]{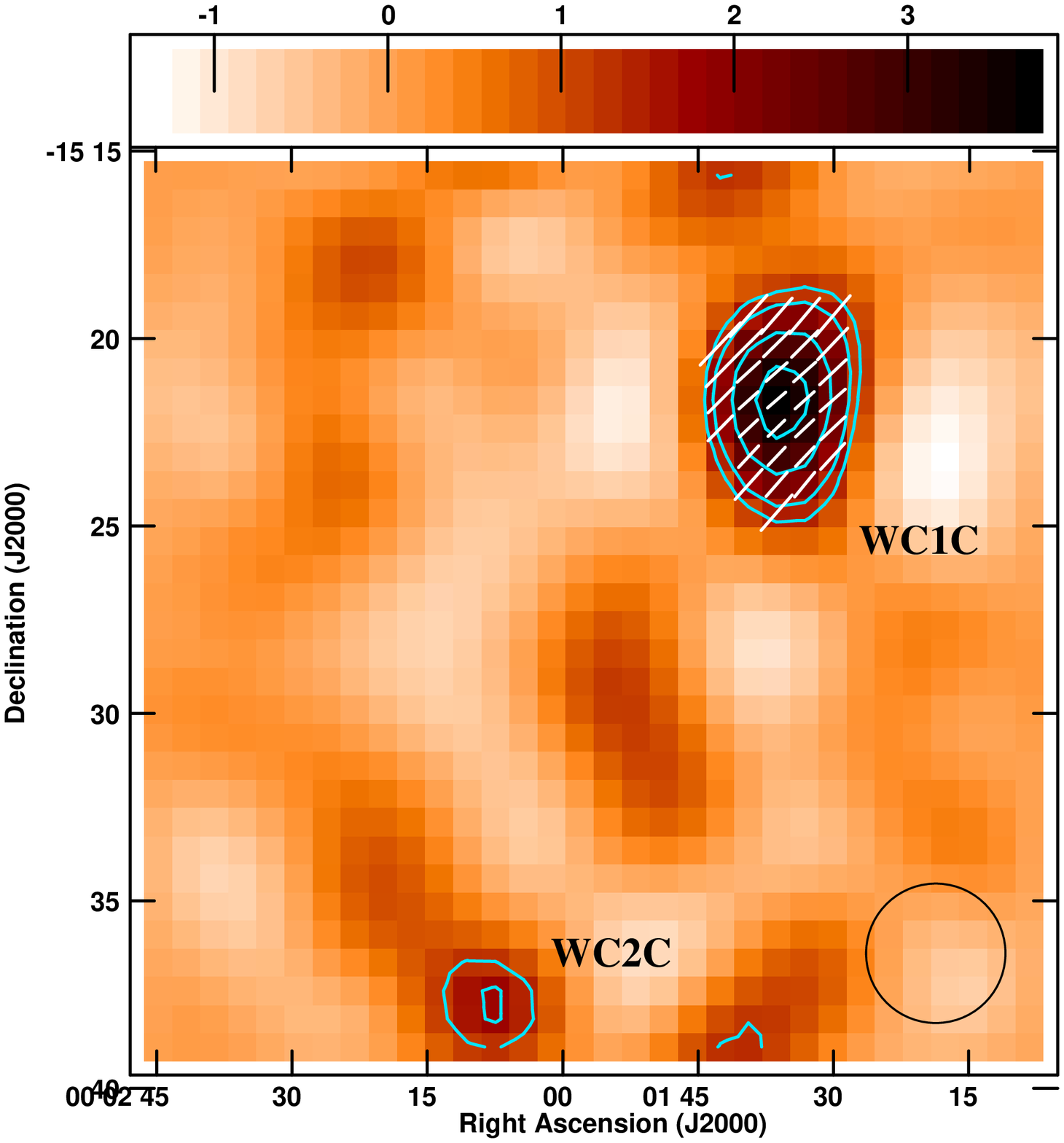}
   \includegraphics[width=8.5cm]{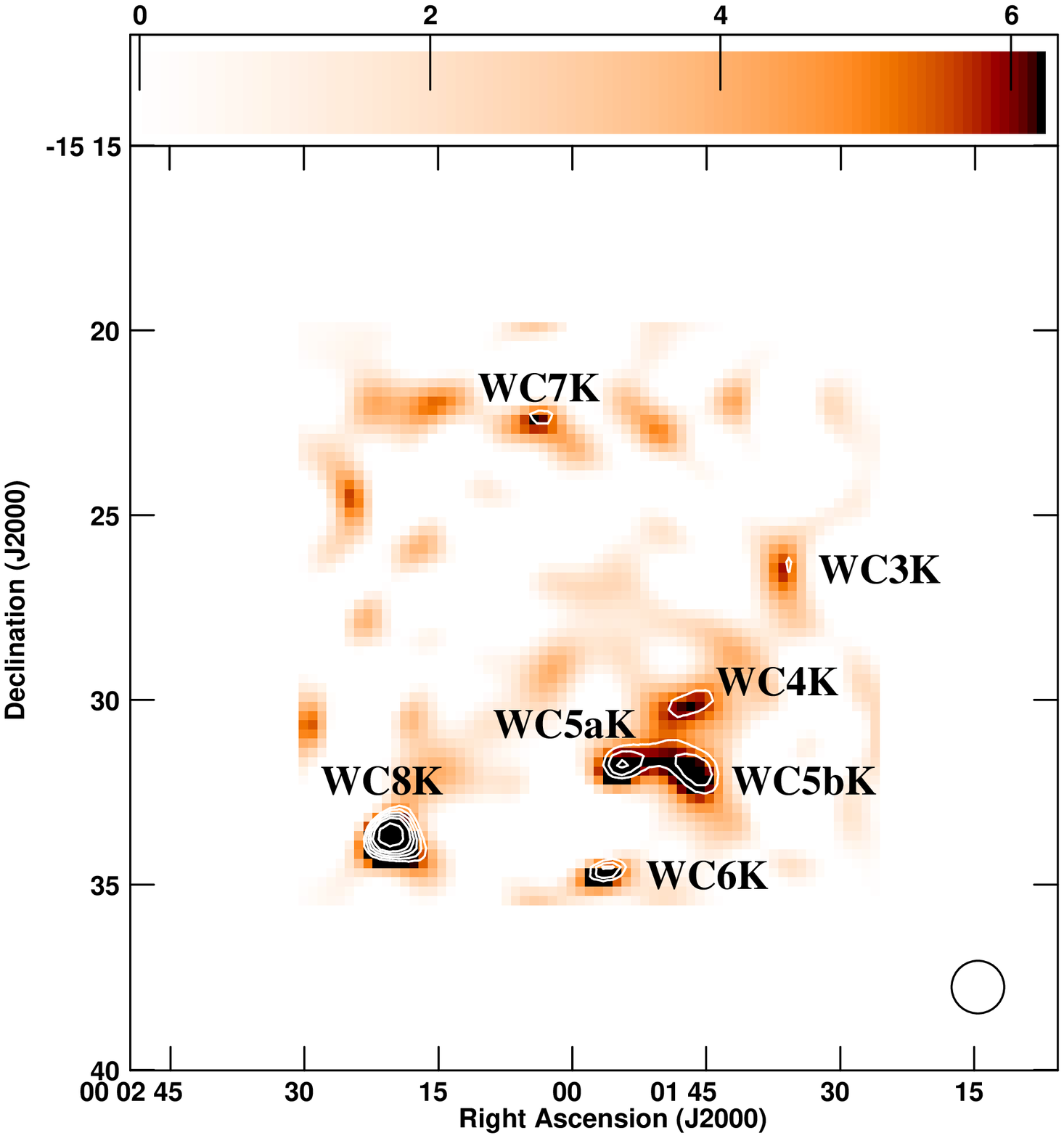}
      \caption{
{\it Left:} SRT C-band total intensity $21^{\prime}\times21^{\prime}$ image of WLM resulting from the spectral average of the inner 80\% of the bandwidth. For this source, the original beam was convolved to a FWHM size of 3.75 arcmin arcmin (bottom-right corner). The noise level is 0.4 mJy/beam. Contour levels are 0.4 $\times$ (3, 4, 6, 8, 10, 10, 15, 20, ..., 80) mJy. Electric field polarization vectors are overlaid. The length of the vectors is proportional to the polarization percentage (with 50\% being a bar of $\sim$ 0.75 arcmin), while their orientation represents the polarization angle. The error on the polarization angle is less than 10\dg, and the fractional polarization is above 3$\sigma_{FPOL}$.  {\it Right:} SRT K-band total intensity $12^{\prime}\times12^{\prime}$ image of WLM resulting from the spectral average of the inner 80\% of the bandwidth. For this source, the original beam was convolved to a FWHM size of 1.4 arcmin (bottom-right corner). The noise level is 1.9 mJy/beam. Contour levels are 1.9 $\times$ (3, 3.5, 4, 4.5, 5, 6, 7) mJy.
              }
         \label{fig:wlmck}
   \end{figure*}

Figures~\ref{fig:n6822ck}, \ref{fig:ic1613ck}, and \ref{fig:wlmck} show the total intensity radio continuum maps of the three target galaxies, NGC\,6822, IC\,1613, and WLM obtained from our SRT observations at C and K band. There is evidence of a number of emitting spots above an average three sigma level in our maps of about 2 and 6 mJy/beam for the C and K bands, respectively. Peak and integrated flux densities are reported in Tables \ref{table:contn6822} to \ref{table:contwlm}. The estimated systemic uncertainty on the absolute flux density calibration for C and K band is $\sim$ 5\% (e.g., \citealt{battistelli2019}) and $\sim$ 15\% (this work)\footnote{This uncertainty at K band has been estimated by applying the same flux calibration performed on the target using 3C~286 to a handful of primary calibrators observed during the observations of NGC\,6822 (in particular, 3C\,48, 3C\,147, and 3C\,295). The measured flux densities have been compared with those obtained using the standard equations and coefficients reported in \citealt{perley2013}. Our estimate is consistent with the value (10\%) reported by \citealt{loru2019}. Our larger uncertainty is likely related to worse average weather conditions and/or the low-elevation of the source(s) during the observations.}, respectively. For unresolved sources, all parameters were determined fitting a 2-dimensional Gaussian with AIPS task JMFIT. In the Gaussian fitting process neither the major nor minor axis sizes were allowed to be smaller than the beam size. For resolved sources, the peak position and flux density were obtained, as before, using JMFIT, while the integrated flux density was estimated using the task TVSTAT over a polygon enclosing the emission down to the 3$\sigma$ level. The errors on the peak positions were directly provided by the task JMFIT. The uncertainty on the peak flux densities has been conservatively assumed to be the noise of the maps, while that of the integrated ones was computed using the formula reported in \citet[their Sect. 3]{panessa2015}. In both cases, when the value obtained was less than the uncertainty on the absolute flux density calibration, the uncertainty was assumed to be 5\% and 15\%, for C and K band, respectively.

In addition, our observation allowed us to obtain Full Stokes information at C-band ($\nu \sim$ 6.7 GHz), and hence, to obtain the percentage of polarization of the radio emission at C band. The left panels of Figures~\ref{fig:n6822ck}, \ref{fig:ic1613ck}, and \ref{fig:wlmck} show electric field polarization vectors overlaid on the C-band total intensity images. The length of the vectors is proportional to the polarization percentage, while their orientation represents the polarization angle. The fractional polarization shown is above 3$\sigma_{FPOL}$. The uncertainty on the fractional polarization ranges from 1 to 20\%, depending on the target and on the position within the map. The error on the polarization angle is less than 10\dg.

As mentioned in Sect.~\ref{sect:obs}, the C and K band cubes for our three galaxies were searched for methanol and water maser emission lines, respectively. At C band, no feature has been found above a 4 sigma level (1$\sigma$ $\approx$ 60 mJy/chan for a 4-km/s wide channel; see also Table~\ref{table:expected}). No water maser detection has been found above a 7 sigma level (1$\sigma$ $\approx$ 100 mJy/chan for a 1.2-km/s wide channel; see also Table~\ref{table:expected}).


\begin{table*}
\caption{Continuum sources detected in NGC\,6822: label, coordinates, peak and integrated flux density, and spectral index between 1.4 and 22 GHz (with $S \propto \nu^{\alpha}$) derived for those sources detected with the SRT at K band having a counterpart in the NVSS (see Sect.~\ref{sect:continuum}).}            
\label{table:contn6822}      
\centering                          
\begin{tabular}{lccccc}        
\hline\hline     
Label   &      RA          &    Dec.  &  Peak & Integrated  & $\alpha^{22}_{1.4}$\\
          &      (J2000)     &  (J2000) &  (mJy/beam)     & (mJy)         &       \\ 
\hline  
 NC1C &  19 44 32.8      $\pm$    0.4  &  -14 42 21     $\pm$     6  &  10.8 $\pm$ 0.6  & 10.8 $\pm$ 0.6 &       \\
 NC2C &  19 44 52.3      $\pm$    0.2   &  -14 43 27     $\pm$     3  &  17.4 $\pm$ 0.9  & 21 $\pm$ 1 &       \\
NC3C &    19 45 06.5      $\pm$    0.9     &  -14 37 31     $\pm$    13  &  4.7 $\pm$ 0.6  &  4.7 $\pm$ 0.6 &       \\
NC4C &    19 45 10.7      $\pm$    0.1     &  -14 51 45     $\pm$     1  &  41 $\pm$ 2  & 48 $\pm$ 2 &       \\
NC5C &     19 45 13.5      $\pm$    0.9    &  -14 57 20     $\pm$    13  &  4.6 $\pm$ 0.6  & 4.6 $\pm$ 0.6  &       \\
NC6C &     19 44 36.9      $\pm$    0.9    & -14 52 04     $\pm$    13  &  4.6 $\pm$ 0.6  & 8 $\pm$ 1 &       \\
NC7C &     19 44 35.0      $\pm$    1.4    & -14 55 24     $\pm$    20  &  3.0 $\pm$ 0.6  &3.0 $\pm$ 0.6  &       \\
\hline
 NC1K &    19 44 33.378      $\pm$    0.005       &   -14 42 18.35     $\pm$     0.07  &  11 $\pm$ 2  & 11  $\pm$ 3 &     0.1 $\pm$ 0.1\\
NC2K &          19 44 52.459      $\pm$    0.006     &   -14 43 14.99     $\pm$     0.09    &    8 $\pm$ 2  &      8 $\pm$ 2    &    -0.3 $\pm$ 0.1   \\
NC4aK &      19 45 07.267      $\pm$    0.007          &   -14 52 42.0     $\pm$     0.1    &    7 $\pm$ 2  &     7 $\pm$ 2      &     -0.6 $\pm$ 0.1  \\
NC4bK &      19 45 13.212      $\pm$    0.007         &  -14 51 26.17     $\pm$     0.09    &   8 $\pm$ 2  &     8 $\pm$ 2     &      -1.1$\pm$ 0.1  \\
 NC6K &     19 44 39.666      $\pm$    0.006     &   -14 51 53.89     $\pm$     0.08  &   9 $\pm$ 2  &    9 $\pm$ 2    &    --   \\
 NC7K &       19 44 39.661      $\pm$    0.005  & -14 56 27.24     $\pm$     0.07 &   10 $\pm$ 2  & 10 $\pm$ 3 &     -- \\
\hline
\end{tabular}
\end{table*}

\begin{table*}
\caption{Continuum sources detected in IC\,1613: label, coordinates, peak and integrated flux density, and spectral index between 1.4 and 22 GHz (with $S \propto \nu^{\alpha}$) derived for those sources detected with the SRT at K band having a counterpart in the NVSS (see Sect.~\ref{sect:continuum}).}            
\label{table:contic1613}      
\centering                          
\begin{tabular}{l c c c c c}        
\hline\hline     
Label   &      RA          &    Dec.  &  Peak & Integrated  & $\alpha^{22}_{1.4}$\\
          &      (J2000)     &  (J2000) &  (mJy/beam)     & (mJy)         &       \\ 
\hline  
 IC1C &   01 04 24.20      $\pm$    0.07     &  02 12 05      $\pm$     1 &    65 $\pm$ 3 &    65 $\pm$ 3  &       \\
 IC2C &   01 04 44.7      $\pm$    0.8    &  02 04 49      $\pm$    12 &  6 $\pm$ 1 &  6 $\pm$ 1  &       \\
 IC3C &    01 05 12.6     $\pm$    1.1    & 02 14 40      $\pm$    17 & 4 $\pm$ 1  & 4 $\pm$ 1  &       \\
\hline
 IC1aK &              01 04 24.836      $\pm$    0.005     &    02 11 48.83      $\pm$     0.08    &   41 $\pm$ 6  & 49 $\pm$ 7   &     -0.12  $\pm$ 0.05 \\
 IC1bK &            01 04 18.35      $\pm$    0.04        &     02 10 03.0      $\pm$     0.5   &     6.3 $\pm$ 1.7  &     6.3 $\pm$ 1.7     &     --    \\
 IC1cK &            01 04 14.97      $\pm$    0.06        &    02 10 55.8     $\pm$     0.8    &    4.8 $\pm$ 1.7   &    4.8 $\pm$ 1.7    &     --   \\
\hline
\end{tabular}
\end{table*}

\begin{table*}
\caption{Continuum sources detected in WLM: label, coordinates, peak and integrated flux density, and spectral index between 1.4 and 22 GHz (with $S \propto \nu^{\alpha}$) derived for those sources detected with the SRT at K band having a counterpart source in the NVSS (see Sect.~\ref{sect:continuum}).}            
\label{table:contwlm}      
\centering                          
\begin{tabular}{l c c c c c}        
\hline\hline     
Label   &      RA          &    Dec.  &  Peak & Integrated  & $\alpha^{22}_{1.4}$\\
          &      (J2000)     &  (J2000) &  (mJy/beam)     & (mJy)         &       \\ 
\hline  
 WC1C &   00 01 35.8      $\pm$    0.8        &   -15 21 37     $\pm$    11  &   4.0 $\pm$ 0.4  &  4.0 $\pm$ 0.6      &       \\
 WC2C &    00 02 08.9     $\pm$    1.8   & -15 37 46    $\pm$    27 &   1.7 $\pm$ 0.4 &  1.7 $\pm$ 0.6  &       \\
\hline
 WC3K &     00 01 36.0      $\pm$    0.8      & -15 26 22     $\pm$    11  &  6 $\pm$ 2    &  6 $\pm$ 2    &   --   \\
 WC4K &     00 01 46.8      $\pm$    0.6      & -15 30 08     $\pm$     9 &  8 $\pm$ 2 &  8 $\pm$ 2  &   --   \\
 WC5aK &     00 01 53.0      $\pm$    0.5     & -15 31 31     $\pm$     8  & 9 $\pm$ 2  & 9 $\pm$ 2 &   --   \\
 WC5bK &     00 01 46.6      $\pm$    0.5     &  -15 31 46     $\pm$     7  &  10 $\pm$ 2  &  10 $\pm$ 2 &  --    \\
WC6K &        00 01 55.8      $\pm$    0.8   & -15 34 51     $\pm$    12  & 6 $\pm$ 2 & 6 $\pm$ 2  &  --    \\
WC7K &    00 02 03.3      $\pm$    0.8      &  -15 22 28     $\pm$    11   &  6 $\pm$ 2  & 6 $\pm$ 2    &  --   \\
WC8K &    00 02 19.8      $\pm$    0.4      & -15 33 43     $\pm$     5  & 13 $\pm$ 2  & 13 $\pm$ 2    &   --  \\
\hline
\end{tabular}
\end{table*}

\section{Discussion}\label{sect:discussion}

\subsection{Radio continuum sources}\label{sect:continuum}

The radio continuum (RC) maps produced by averaging in frequency the radio data cubes show the presence of several radio continuum sources both at C and K band. As discussed in two past works by \citet{chyzy2003} and \citet{chyzy2011}, and, more recently, by \citet{hindson2018}, some of these sources may be unrelated to the target objects, being emission produced by background sources. In order to try evaluating which sources belong to the target galaxies, we have derived their spectral index, considered polarization properties and compared our radio continuum maps with those taken at other frequencies, e.g. radio, optical and, UV. 

In order to derive the spectral indices of the sources, we have compared our SRT K-band maps with those taken from the NRAO VLA Sky Survey (NVSS) at 1.4 GHz (\citealt{condon1998}). The resolution of the two maps are comparable, 50$^\prime$ and 45$^\prime$ for the SRT and VLA images, respectively, allowing to identify the main centers of radio continuum emission at both frequencies. The last columns of Tables \ref{table:contn6822}, \ref{table:contic1613}, and \ref{table:contwlm} report the spectral index between 1.4 and 22 GHz, $\alpha^{22}_{1.4}$ (with $S \propto \nu^{\alpha}$) of the sources detected above a three sigma level, at both frequencies. The error on the spectral indices has been computed, using the standard error propagation formula, from the errors reported for the sources at K-band (Tables \ref{table:contn6822} to \ref{table:contwlm}) and those obtained from the NVSS source catalog (\citealt{condon1998}).

   \begin{figure}
    \includegraphics[width=10.0cm]{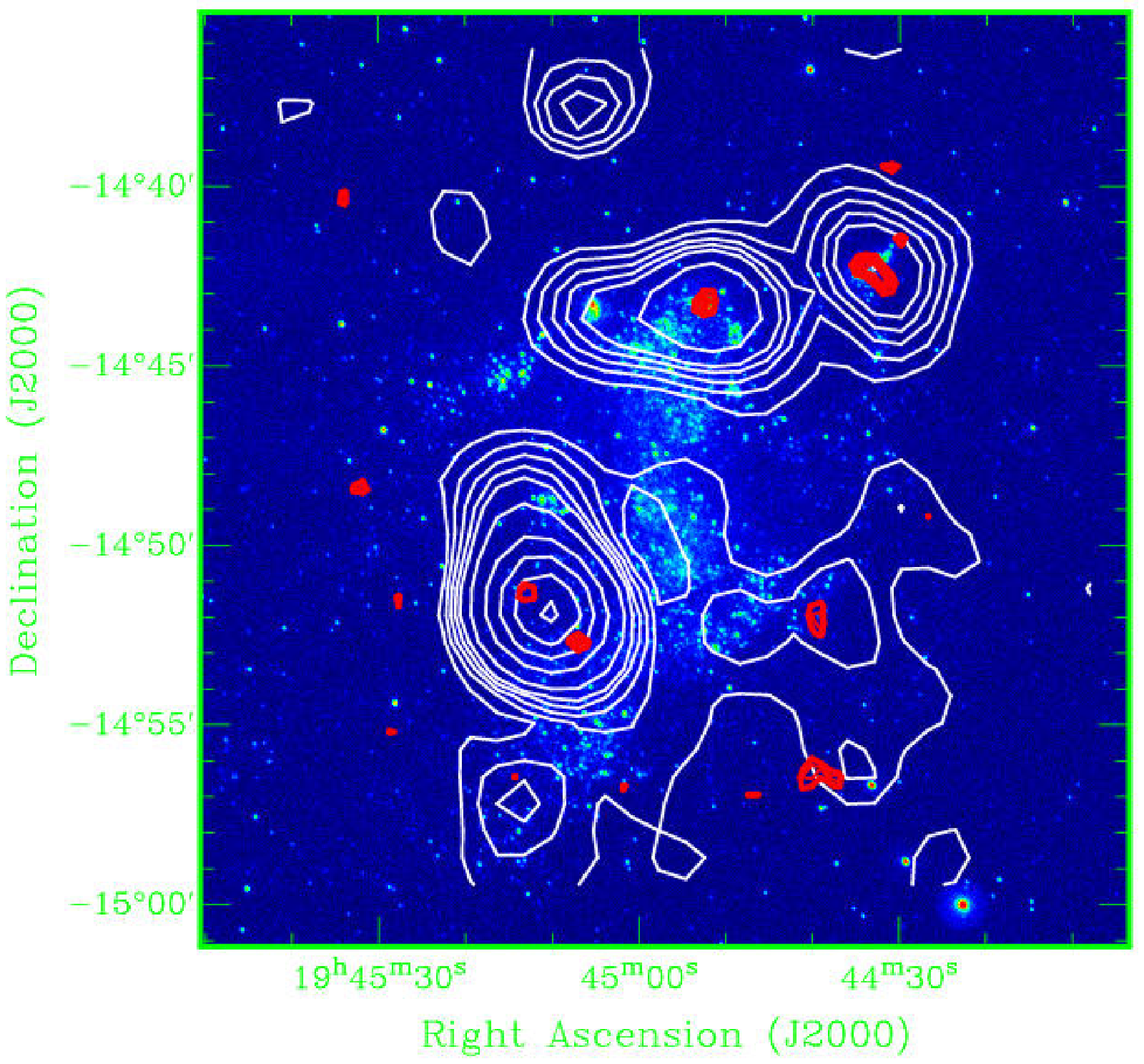}
      \caption{Combination of a FUV Galex map of NGC\,6822 with the SRT radio continuum C (white) and K (red) band contour maps overlaid. Contour levels are 0.65 $\times$ (2, 4, 6, 8, 10, 15, 25, 35, 45, 55) mJy and 2 $\times$ (3, 3.5, 4, 4.5, 5) mJy, for the C and K band, respectively.
}
         \label{fig:combo_n6822}
   \end{figure}

   \begin{figure}
     \includegraphics[width=8.5cm]{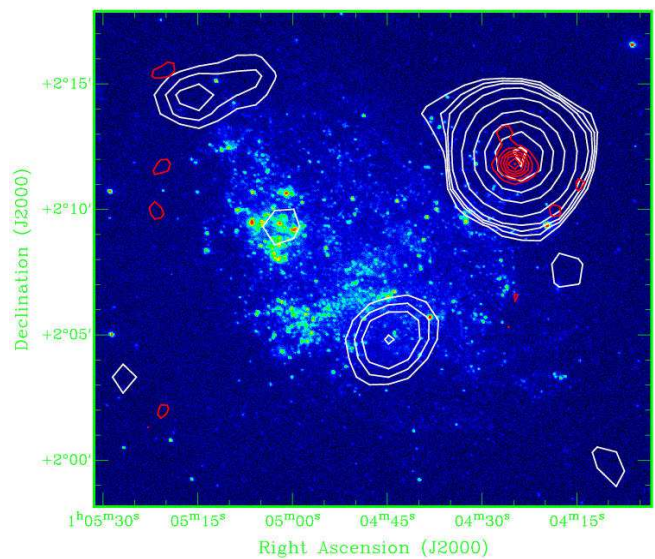}
      \caption{Combination of a FUV Galex map of IC\,1613 with the SRT radio continuum C (white) and K (red) band contour maps overlaid. Contour levels are 1.0 $\times$ (2, 3, 4, 6, 12, 24, 36, 48, 60) mJy and 1.7 $\times$ (3, 6, 9, 12, 15, 18, 21) mJy, for the C and K band, respectively.
}
         \label{fig:combo_ic1613}
   \end{figure}

   \begin{figure}
    \includegraphics[width=9.0cm]{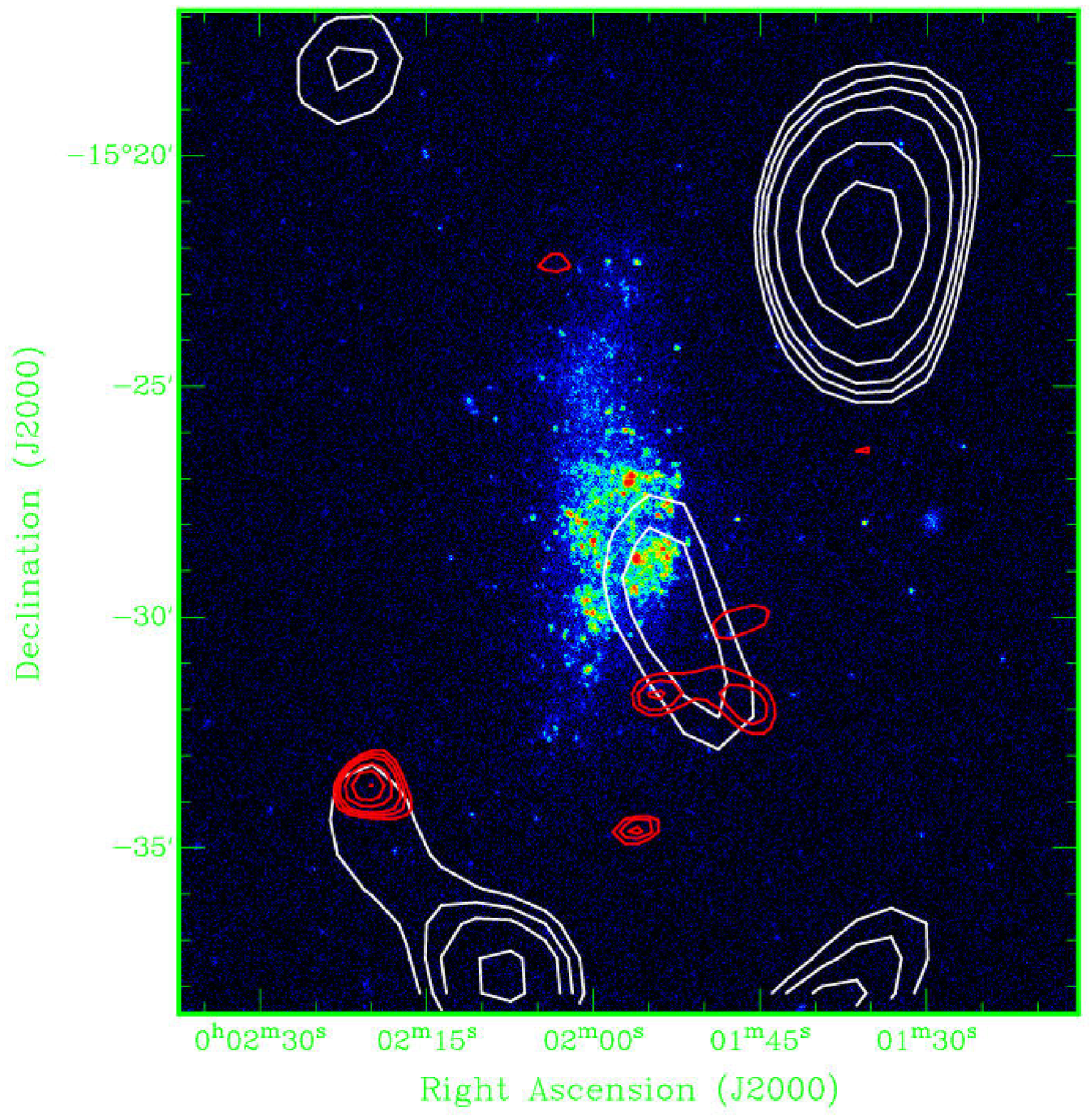}
      \caption{Combination of a FUV Galex map of WLM with the SRT radio continuum C (white) and K (red) band contour maps overlaid. Contour levels are 0.4 $\times$ (2, 2.5, 3, 4, 6, 8) mJy and 1.9 $\times$ (3, 3.5, 4, 5, 6, 7) mJy, for the C and K band, respectively.
}
         \label{fig:combo_wlm}
   \end{figure}

A preliminary association of the radio continuum sources with the galaxies is mainly based on the presence (or absence) of optical/UV/FUV counterparts (Figs.~\ref{fig:combo_n6822}, ~\ref{fig:combo_ic1613}, \& \ref{fig:combo_wlm}) . In NGC6822, the target labeled NC3C is clearly offset w.r.t. the optical body of the galaxy. This and the presence of linear polarization (up to $\sim$20$\pm$5\%) hints at a non-thermal nature of the emission. This makes the object a background source candidate. Similarly, background source candidates are also sources IC1C in IC1613, and WC1C, WC2C, and WC8K in WLM. The other radio continuum sources detected in the three targets lie to some extent within the optical body of the galaxies. However, for NGC6822, NC4C shows linear polarization in the northeastern extension with a percentage between 10$\pm$5 and 22$\pm$10, and also, although at a lower level (5$\pm$1\%), close to the source peak. Indeed, NC4C is also resolved into, at least, two sources at L- and K-bands, both with negative spectral index (see Table~\ref{table:contn6822}), thus hinting at a nonthermal nature of the emission. This, together with the lack of evident counterparts at other wavelengths, conceivably makes NC4C a background object. The same conclusion was also reached by \citet{chyzy2003}.Notably, at resolution and frequencies comparable to those in our study, polarization fractions of the order of some 10 percent (as also found in NC3C, see above) are not unusual in radio sources, and, for example, in the case of tailed radio galaxies, are expected to increase along the tail reaching values above 50\% (see, e.g., \citealt{feretti1998}). All other sources in NGC6822 are either related to the target objects or are of ambiguous nature\footnote{We define as ambiguous those sources for which it is not possible, from the information available to date, to confidently determine or exclude their association to the galaxies}. In the case of IC1613, the detected radio continuum sources are all of ambiguous nature. Indeed, the case of IC1C is quite peculiar. The source is linearly polarized (between 5 and 10 \%) and is resolved into three centers of emission at K band. However, the spectral index of the brightest component is flat, possibly indicating a thermal origin. Thus, despite the source being at the outskirts of the optical body of the galaxy, it cannot be confidently classified as a background source, although in the following discussion it will be treated as such. For WLM, the spatial distance of the radio continuum sources to the main optical emitting regions favors a scenario in which these sources are unrelated to the galaxy, with the exception of sources WC4K, and WC5K (a\&b), that have some (weak) optical counterparts.

\begin{table*}
\caption{Galaxy name, integrated 6.7 GHz flux and luminosity (upper line: including all sources; lower line: excluding background sources)  and corresponding 1.4 GHz flux and luminosity (derived assuming a spectral index of 0.75; $S \propto \nu^{\alpha}$), SFRs estimated using Eq.~\ref{eq:sfr1}, Eq.~\ref{eq:sfr2}, and from extinction-corrected H$\alpha$ fluxes (from \citealt{mateo1998}). For details, see also Sect.~\ref{sect:continuum}.}
\label{table:sfr}
\begin{tabular}{lccccccc}
                &       &    & & & &   & \\
\hline
\\
Source          &   S$^{6.7-GHz}_{RC}$ & L$^{6.7-GHz}_{RC}$  & S$^{1.4-GHz}_{RC}$ & L$^{1.4-GHz}_{RC}$  & SFR$_{RC1}^{a)}$  &  SFR$_{RC1}^{b)}$ & SFR$_{H\alpha}^{c)}$ \\
                &   (mJy/beam)      & (Watt Hz$^{-1}$)  & (mJy/beam)      &  (Watt Hz$^{-1}$) & (\solmass yr$^{-1}$)       &  (\solmass yr$^{-1}$) \\
\hline
NGC\,6822       &      96$\pm$5           &   (2.8$\pm$0.5)$\times$10$^{18}$ &      312$\pm$16         &   (9$\pm$2)$\times$10$^{18}$               &    0.014$\pm$0.003              &   0.011$\pm$0.002    & 0.06         \\
                &      44$\pm$5           &   (1.4$\pm$0.3)$\times$10$^{18}$ &      141$\pm$16         &   (4.0$\pm$0.8)$\times$10$^{18}$               &    0.006$\pm$0.001                &   0.005$\pm$0.001     &            \\
 IC\,1613       &       81$\pm$4           &  (4.8$\pm$0.5)$\times$10$^{18}$  &     263$\pm$13           &   (1.5$\pm$0.2)$\times$10$^{19}$               &    0.026$\pm$0.003                &   0.018$\pm$0.002    & 0.003                \\
                &      16$\pm$5           &  (1.0$\pm$0.3)$\times$10$^{18}$  &      53$\pm$16          &    (3$\pm$1)$\times$10$^{18}$             &    0.004$\pm$0.002                &    0.004$\pm$0.001    &            \\  
WLM            &       9$\pm$1         &   (9$\pm$1)$\times$10$^{17}$ &        28$\pm$1           &  (2.9$\pm$0.3)$\times$10$^{18}$               &    0.0041$\pm$0.0007                &   0.0034$\pm$0.0004      & 0.003             \\
                &        3$\pm$1         &   (1$\pm$1)$\times$10$^{17}$ &      10 $\pm$2          &     (10$\pm$3)$\times$10$^{17}$              &    0.0012$\pm$0.0006                &     0.0012$\pm$0.0003    &             \\

\hline
\end{tabular}
%
\\
$^{a)}$ from Eq.~\ref{eq:sfr1}\\
$^{b)}$ from Eq.~\ref{eq:sfr2}\\
$^{c)}$ from Mateo (1998)\\
\end{table*}

In Table~\ref{table:sfr}, we report the C-band radio continuum flux of the target galaxies derived by integrating over an area covering the optical body of the galaxy. Integrated fluxes were measured both including the emission coming from all sources in the map and subtracting the contribution by the most likely background source candidates (see the discussion above)\footnote{For NGC\,6822, we subtracted the contribution of sources NC3C and NC4C. For IC\,1613, we removed the flux pertaining to IC1C. For WLM, both radio continuum sources, WC1C and WC2C were excluded.}. Uncertainties on the integrated fluxes were computed using the formula reported in \citet[their Sect. 3]{panessa2015}. In all cases, the values obtained were larger than the uncertainty on the absolute flux density calibration, i.e. 5\% (see Sect.~\ref{sect:results}). The star formation rate (SFR) of each galaxy from {the emission} at 6.7 GHz, computed by using the following RC-SFR relation, taken from \citet[their Eq. 10]{hindson2018}, is also reported:

\begin{equation}\label{eq:sfr1}
\frac{L_{RC}}{W\,Hz^{-1}} = 10^{A} \, \bigg(\frac{SFR}{\solmass\,yr^{-1}}\bigg)^n
\end{equation}

where {\it A} and {\it n} are the Best-fit parameters for the RC-SFR relation, derived by Hindson et al. for their data using different approaches (their Section 4.6). For our estimates, we used an average value of these parameters (\citet[their Table 6]{hindson2018}), of 20.1 and 0.9, for {\it A} and {\it n}, respectively. 

In Table~\ref{table:sfr}, we also report the computed values of the SFR in the three galaxies, obtained from the 1.4 GHz luminosity through the equation by \citet[and references therein; their Eq.~2]{hopkins2002}:

\begin{equation}\label{eq:sfr2}
SFR = 5.5 \times \bigg(\frac{L_{1.4-GHz}}{4.6 \times 10^{21}\:W\,Hz^{-1}}\bigg)\:\solmass\,yr^{-1}
\end{equation}

The 1.4 GHz luminosities are derived from the 6.7 GHz luminosities, assuming a spectral index of 0.75 (\citealt{gioia1982}).\footnote{This value is the average spectral index value estimated for spiral galaxies. For dwarf irregular galaxies, flatter radio continuum spectra than those of 'normal' spiral galaxies, have indeed also been suggested (e.g., \citealt{klein1986}). This would, however, reduce the extrapolated 1.4-GHz luminosities, and the corresponding SFRs, thus further reinforcing our considerations.} 

The last column of Table~\ref{table:sfr} reports the SFR of the three targets reported by Mateo (1998) and derived by using the H$\alpha$ emission. 

Reported errors for the luminosities and SFRs take into account, through the standard errors propagation equation, both the flux density uncertainties and those for the galaxy distance (see Table~\ref{table:targets}). When comparing the SFRs obtained in this work with those reported, without an associated uncertainty, by Mateo (1998), it has, however, to be considered that the distances used for the galaxies are the same.

The SFRs derived at radio frequencies from Eqs.~\ref{eq:sfr1} and \ref{eq:sfr2} agree fairly well. For NGC\,6822 all RC-based SFRs are (significantly) lower than that obtained with H$\alpha$. For IC\,1613, instead, excluding the RC emission background source candidates in the SFR computation provides a better match with the H$\alpha$-based SFR, possibly confirming that IC1C is indeed not associated with the galaxy. For WLM, SFRs derived from the radio continuum are consistent with those derived by using the H$\alpha$ emission only when considering all the emission present in the field as belonging to the galaxy (that, as said before, is unlikely). 
Overall, however, for two out of our three targets, the global SFRs derived by 'optical means' are higher than those derived by us using radio continuum data. This is consistent with the conclusions of Hindson et al. (2018). Indeed, they find that their targets (i.e., dwarf galaxies) follow the theoretical predictions of the RC-SFR relation only in individual regions of enhanced radio continuum emission but, apparently, not when considering the entire optical disks of the galaxies.

\subsection{The paucity of methanol and water masers in dwarf galaxies}\label{sect:search}

For this project, for the first time, the full optical body of the galaxies NGC\,6822, IC\,1613, and WLM  was searched for spectral line emission at 6.7 and 22 GHz. In the past, only water maser emission was searched for in these galaxies but these searches were limited to individual regions, typically associated with enhanced H$\alpha$ or FIR emission \citep[see][hereafter BHB, and references therein]{brunthaler2006}. As mentioned in Sect.~\ref{sect:intro}, the possible absence of an optical counterpart associated with a maser source, makes the need to create fully-sampled spectroscopic images of the entire optical extension of the galaxies particularly important, in order to confidently assess the occurrence of maser emission in the target galaxy.
   
For the three targets of our study, the lack of confident line detections, both at C and K bands, indicates a general absence of (strong) maser emission in dwarf galaxies. In the following two subsections, we try to associate the aforementioned absence to the main peculiarities of dwarf galaxies.

\subsubsection{The modest star forming rate}\label{sect:reason1}
For water masers, our result confirms the conclusions obtained by BHB. Indeed, for water masers, their study indicates that a value for the expected number (N) of masers can be computed using the luminosity function (LF) for H$_{2}$O masers in our Galaxy, empirically derived by \citet{greenhill1990} and accounting for the different star formation rates (SFR) of the target galaxy and that of the Milky Way\footnote{The ratio is based on the assumption that the rates of maser production and star formation are roughly proportional. This assumption is plausible when, as in this case, we are considering exclusively masers associated with star formation. Indeed, there is no AGN related H$_{2}$O (mega)maser in the LG.}. For H$_{2}$O masers, the expected number of sources with an isotropic luminosity larger than a certain value, $L_{H_{2}O}$, is:    
\begin{equation}\label{eq:h2oexpnum}
N^{H_{2}O}_{\rm expected} = 10^{\{-0.6 \times [1 + log (L_{H_{2}O})]\}} \times (SFR_{\rm target} / SFR_{\rm Milky Way}),
\end{equation}
where the luminosity in the 22 GHz line $L_{H_{2}O}$ (in \solum) is taken as the (4$\sigma$) detection threshold of our observations, $S$ (in Jy), derived from the maser linewidth assumed to be equal to the velocity width of the channel (${\Delta}{v}$; 1.2 km/s, in our case), and from the distance (in Mpc) of our target galaxy, using the equation 
\begin{equation}\label{eq:maslumwater}
L_{H_{2}O} = 0.023 \times S \times {\Delta}{v} \times D^{2}. 
\end{equation}

For example, with a 3$\sigma$ detection threshold of 30 mJy/chan (millyJansky/channel) for a 1-km/s wide channel, and using SFRs for the Milky Way and M\,31 of 4 \solmass yr$^{-1}$ (Diehl et al.\ 2006; from the frequency of core collapse supernovae derived by gamma-ray measurements) and 0.35 \solmass yr$^{-1}$ (\citealt{walterbos1994}; from the extinction-corrected H$\alpha$ luminosity), respectively, we obtain an expected number of maser sources in M\,31 of about 3. This number is fully consistent with the water maser detections obtained in M\,31 by \citet{darling2011} and \citet{darling2016}. Similarly consistent results are also found for the other LG galaxies where water masers have been found\footnote{Slightly different numbers and conclusions are, however, reached for the Large and Small Magellanic Clouds by \citet{breen2013}}, with the exception of IC\,10 in which an overabundance of masers with respect to the expected number is detected (see BHB, and references therein). Table~\ref{table:expected} reports the number of water masers expected for the three galaxies in our study, with a luminosity larger than the threshold corresponding to the 4$\sigma$ level of our cubes. As evident, the expected number is small for all galaxies (close to zero for IC\,1613 and WLM). Indeed, for NGC\,6822, lowering the luminosity threshold by an order of magnitude (2-3 orders of magnitude are instead required for dwarfs with SFRs similar to those of IC\,1613 and WLM) would bring the expected number to unity, opening the possibility for a detection but this would require an extremely long observation, when not led with the aid of a multi-feed receiver or with an extremely sensitive facility. 

\begin{table*}
\caption{Galaxy name (col. [1]); average 1$\sigma$ $rms$ noise levels in the C-band and K-band cubes (cols. [2] and [6]); methanol and water maser luminosity detection limits (cols. [3] and [7]), computed from Eqs.~\ref{eq:maslummeth} and .~\ref{eq:maslumwater}, respectively, using the 4$\sigma$ $rms$ levels in the cubes and a channel width of 4 and 1.2 \kms, respectively; number of expected  methanol and water masers (cols. [4] and [8]), derived from Eqs.~\ref{eq:methexpnum} and \ref{eq:h2oexpnum}, respectively; number of methanol and water masers detected (cols. [5] and [9]).}
\label{table:expected}
\begin{tabular}{lcccccccc}
                &               &                  &                &                     &                    &                      &                     \\
\hline
\\
Source          & $rms_{CH{_3}OH}$ & L$_{CH{_3}OH}$      & N$_{CH{_3}OH}^{exp,\,a)}$ & N$_{CH{_3}OH}^{det}$ &    $rms_{H{_2}O}$ & L$_{H{_2}O}$  & N$_{H{_2}O}^{exp,\,a)}$ & N$_{H{_2}O}^{det}$ \\
                &    (mJy)      & (\solum)         &                       &                      & (mJy)          & (\solum)     &                    &                      \\
\hline
NGC\,6822       &  60 &                  1.6  $\times 10^{-3}$ &   0.1     &   0       &         80 & 2.2 $\times10^{-3}$   & 0.1  & 0              \\ 
 IC\,1613       &        60            &      3.3  $\times 10^{-3}$              &   0.004    &     0           &70  & 3.9 $\times 10^{-3}$                & 0.005  &  0  \\
 WLM            &    65                &    6.3  $\times 10^{-3}$                &   0.003  &     0             & 160           & 15.5 $\times 10^{-3}$ & 0.002       &  0 \\
\hline
\end{tabular}
%
\\
$^{a)}$ The expected maser numbers are computed using the SFRs derived from extinction-corrected H$\alpha$ fluxes, as reported in Mateo (1998; see also Table~\ref{table:sfr})\\
\end{table*}

For methanol masers, a relation similar to Eq.~\ref{eq:h2oexpnum} can also be derived by using the information reported in \citet{quiroganunez2017}. Using their value for the slope of the LF ($\alpha$ = -1.43) and the reported total number of Galactic methanol masers of 1300 (\citealt{quiroganunez2017}), computed integrating between $10^{-8}$ and $10^{-3}$ \solum\ (from \citealt{pestalozzi2007}), the complete relation for the LF can be now be expressed as $f(L) = 0.2 \times L^{-1.43}$. Hence, the expected number of 6.7-GHz methanol masers with an isotropic luminosity larger than a certain value, $L_{CH_{3}OH}$, is:    
\begin{equation}\label{eq:methexpnum}
N^{CH_{3}OH}_{\rm expected} = 10^{\{-0.4 \times [0.8 + log (L_{CH_{3}OH})]\}} \times (SFR_{\rm target} / SFR_{\rm Milky Way}),
\end{equation}
where $L_{CH_{3}OH}$ (in \solum) is taken as the (4$\sigma$) detection threshold of our observation, $S$ (in Jy), from the maser linewidth assumed equal to the velocity width of the channel (${\Delta}{v}$; 4.1 km/s, in our case), and from the distance (in Mpc) to our target galaxy, using the equation
\begin{equation}\label{eq:maslummeth}
L_{CH_{3}OH} = 0.0069 \times S \times {\Delta}{v} \times D^{2}. 
\end{equation}

Table~\ref{table:expected} reports the number of methanol masers expected for the three galaxies of our study, with a luminosity larger than the threshold corresponding to the 4$\sigma$ level of our cubes. Also in this case, the expected numbers are small, well below unity.

For the computation of the expected number of both water and methanol masers, we used the SFRs derived from extinction-corrected H$\alpha$ fluxes, as reported in \citep[see also Table~\ref{table:sfr}]{mateo1998}. Using the SFRs derived by us from the radio continuum (Table~\ref{table:sfr}) would yield even lower values for the expected numbers and would not change our conclusions.

\subsubsection{The low metallicity}\label{sect:reason2}

Dwarf irregular galaxies are relatively low mass, gas-rich, metal-poor, and are sometimes forming stars as shown by their \hii\ regions (e.g., \citealt{lee2005}). Indeed, our three targets have low metallicities, with $12 + log(O/H)$ values of 8.0, 7.9, and 7.8, for NGC\,6822 (\citealt{Schruba2017}), IC\,1613 (\citealt{bresolin2007}), and WLM (\citealt{rubio2015}), respectively. This corresponds approximately to 15-20 \% of the solar value ($\sim$ 8.7, from \citealt{asplund2009}). While, as discussed in the previous section, the star formation rates for these galaxies are already lower than those of regular spirals, resulting in low expected numbers of masers in dwarf galaxies, it is reasonable to assume that the low oxygen and carbon abundances, together with a low abundance of other processed elements and dust, may influence the possibility to have a large fraction of water and methanol molecules, further reducing the possibility to have maser emission from these species. The chemical modelling of dark clouds in the Magellanic clouds by \citet{millar1990} indicates, however, that the scaling between molecular and elemental abundances, except for the CO molecule, does not follow a simple pattern and that the molecular abundances depend on the varying C, O and N elemental abundances in the models in a complex manner. From the observational point of view, the CO content and distribution in low-metallicity dwarf galaxies has been recently studied using ALMA observations by, e.g., \citet{rubio2015} and \citet{Schruba2017}. Their conclusions mainly point to the presence of compact CO clumps, with small radii, narrow line width, and low filling factor across these galaxies. The confined nature of the CO emission agrees with a scenario invoking a lack of dust shielding that would push CO emission deep into molecular clouds in low metallicity objects. Within this framework, the same absence of dust shielding, and, correspondingly, smaller fraction of dust grains could prevent the formation of some molecules, as methanol and water are typically formed via reactions on grains surfaces. It could also increase the efficiency of photo-dissociation and photo-ionization on the molecules after their formation caused by external radiation. This seems to be true for the methanol molecule, thought to be produced by the hydrogenation of CO on dust grains and released into the gas phase by thermal and/or non-thermal desorption (e.g., \citealt{watanabe2002}). Indeed, a low abundance of dust grains would then make the methanol formation inefficient, explaining the lack of detected methanol lines in the LMC and IC\,10 (\citealt{nishimura2016}, and references therein). Hence, it is not far-fetched to assume that a similar scenario applies also to the water molecule, although this hypothesis has not, to our knowledge, been investigated yet. If this scenario is confirmed, the expected number of maser sources suggested by Eqs.~\ref{eq:h2oexpnum} and \ref{eq:methexpnum} would provide upper limits.


\section{Conclusions}\label{sect:conclusions}

During Spring 2016, we obtained on-the-fly (single-feed) maps of the continuum and spectral line radio emission over the full optical extents of 14 Local Group dwarf galaxies, belonging to two highly-symmetric dynamical planar structures, with the Sardinia radio Telescope (SRT), at C and K band. The data for the first three galaxies, NGC\,6822, IC\,1613, and WLM, were reduced and analyzed in order to obtain the distribution and intensity of the radio continuum emission from the targets and, to search, for the first time across the entire extension of the galaxies, for methanol and water (maser) sources to be used for follow-up VLBI proper motion studies.  
We summarize our results and conclusions as follows:

\begin{itemize}
\item radio continuum emission has been detected in all maps, mainly coming from discrete sources, although a diffuse component is also seen in the C-band map of NGC\,6822. Spectral indices for the sources have been estimated with values consistent with a mix of thermal and non-thermal origin. Polarization properties at C band have been also inferred, showing that some sources are linearly polarized with percentages higher than 20 \%. Indeed, some of the radio continuum sources have been found unrelated to the galaxies and are likely produced by background objects. 
\item global star formation rates have been derived from the radio continuum emission detected from the galaxies. They have been found to be, on average, lower than those obtained at other wavelengths. This may suggest that, as also concluded by other authors, dwarf galaxies do not follow the theoretical predictions of the RC-SFR relation when the whole optical disks are considered, while the correlation holds for individual regions of enhanced radio continuum emission.
\item No 6.7-GHz methanol and 22-GHz water maser emission has been detected in the three targets, with isotropic luminosity above a few thousandths of solar luminosities. This is consistent with past studies that indicate that the number of maser sources (associated to star-formation activity) in a galaxy should decrease with its star formation rate. In addition, a low metallicity may reduce the abundance of the masing molecular species, because of a smaller fraction of dust grains where such molecules typically form and from which molecules are shielded by the strong radiation field. Hence, the combination of low star formation rate and low abundance of heavy elements in irregular dwarf galaxies could work against a widespread presence of maser sources in this class of galaxies. 
 \end{itemize}
 
Overall, our study has further shown the potential of the SRT in performing on-the-fly fully-sampled spectropolarimetric maps of extended sources in the sky. It has, however, also pointed out the necessity to perform sufficiently sensitive measurements (between 1 to 3 orders of magnitude deeper, depending on the target's star formation rate, than the one reported in this work) when weaker or elusive star formation related (maser) lines are searched for in dwarf galaxies. This necessarily implies extremely long integration times, especially at high frequencies, something unsustainable for modern, often over-subscribed facilities. Therefore, the use of multi-feed receivers in this kind of studies becomes mandatory. Indeed, a 7-feed K-band receiver has been recently offered also for spectropolarimetric measurements at the SRT, making deep spectral line surveys in extended objects now possible. In addition, the SKA telescope and the ng-VLA, with their unique sensitivity, will ultimately be game-changers in this kind of studies. 

\section*{Acknowledgements}

The SRT ESP activities were made possible thanks to the invaluable support of the entire SRT Operations Team. In particular, we would like to thank Ettore Carretti for all the work done as Officer-in Charge of the SRT during the ESP, and Franco Buffa for promptly providing us with the relevant weather information during observations. We thank the anonymous referee for his/her comments. We are grateful to Federica Govoni for useful discussion, and to Luis Quiroga-Nu\~nez for sharing information on the Galactic methanol maser luminosity function. The Sardinia Radio Telescope is funded by the Department of University and Research (MIUR), Italian Space Agency (ASI), and the Autonomous Region of Sardinia (RAS) and is operated as National Facility by the National Institute for Astrophysics (INAF). The development of the SARDARA backend has been funded by the Autonomous Region of Sardinia (RAS) using resources from the Regional Law 7/2007 ``Promotion of the scientific research and technological innovation in Sardinia'' in the context of the research project CRP-49231 (year 2011, PI Possenti): ``High resolution sampling of the Universe in the radio band: an unprecedented instrument to understand the fundamental laws of the nature''. This publication is partly based on observations with the 100-m telescope of the MPIfR (Max-Planck-Institut f\"{u}r Radioastronomie) at Effelsberg. This research has made use of the NASA/IPAC Extragalactic Database (NED), which is operated by the Jet Propulsion Laboratory, California Institute of Technology, under contract with the National Aeronautics and Space Administration.





\begin{thebibliography}{99}
\bibitem[\protect\citeauthoryear{Amiri \& Darling}{2016}]{amiri2016}Amiri, N. \& Darling, J. 2016, \apj, 826, 136A
\bibitem[\protect\citeauthoryear{Argon et al.}{1994}]{argon1994}Argon, A. L., Greenhill, L. J., Moran, J. M., et al. 1994, \apj, 422, 586A
\bibitem[\protect\citeauthoryear{Asplund et al.}{2009}]{asplund2009}Asplund, M., Grevesse, N., Sauval, A. J., Scott, P. 2009, \araa, 47, 481A
\bibitem[\protect\citeauthoryear{Baan \& Haschick}{1994}]{baan1994}Baan, W. A. \& Haschick, A. 1994, \apj, 424L, 33B
\bibitem[\protect\citeauthoryear{Battistelli et al.}{2019}]{battistelli2019} Battistelli, E. S., Fatigoni, S., Murgia, M., et al. 2019, \apj, 877L, 31B 
\bibitem[\protect\citeauthoryear{Becker et al.}{1993}]{becker1993}Becker, R., Henkel, C., Wilson, T. L., Wouterloot, J. G. A. 1993, \aap, 268, 483B
\bibitem[\protect\citeauthoryear{Bolli et al.}{2015}]{bolli2015}Bolli, P., Orlati, A.,  Stringhetti, L., et al. 2015, JAI, 450008B
\bibitem[\protect\citeauthoryear{Bresolin et al.}{2007}]{bresolin2007}Bresolin, F., Urbaneja, M. A., Gieren, W., Pietrzy\'nski, G., Kudritzki, R. 2007, \apj, 671, 2028B
\bibitem[\protect\citeauthoryear{Breen et al.}{2013}]{breen2013}Breen, S. L., Lovell, J. E. J., Ellingsen, S. P., et al. 2013, \mnras, 432, 1382B
\bibitem[\protect\citeauthoryear{Brunthaler et al.}{2005}]{brunthaler2005}Brunthaler, A., Reid, M. J., Falcke, H., Greenhill, L. J., Henkel, C. 2005, \sci, 307, 1440B
\bibitem[\protect\citeauthoryear{Brunthaler et al.}{2006}]{brunthaler2006}Brunthaler, A., Henkel, C., de Blok, W. J. G., et al. 2006, \aap, 457, 109B; BHB
\bibitem[\protect\citeauthoryear{Brunthaler et al.}{2007}]{brunthaler2007}Brunthaler, A., Reid, M. J., Falcke, H., Henkel, C., Menten, K. M. 2007, \aap, 462, 101B
\bibitem[\protect\citeauthoryear{Buck et al.}{2016}]{buck2016}Buck, T., Dutton, A. A., Macci\`{o}, Andrea, V. 2016, \mnras, 460, 4348B
\bibitem[\protect\citeauthoryear{Chy\.zy et al.}{2003}]{chyzy2003}Chy\.zy, K. T., Knapik, J., Bomans, D. J., et al. 2003, \aap, 405, 513C 
\bibitem[\protect\citeauthoryear{Chy\.zy et al.}{2011}]{chyzy2011}Chy\.zy, K. T., We\.zgowiec, M., Beck, R., Bomans, D. J. 2011, \aap, 529A, 94C
\bibitem[\protect\citeauthoryear{Condon et al.}{1998}]{condon1998}Condon, J. J., Cotton, W. D., Greisen, E. W., et al. 1998, \aj, 115, 1693
\bibitem[\protect\citeauthoryear{Darling}{2011}]{darling2011}Darling, J. 2011, \apj, 732L, 2D
\bibitem[\protect\citeauthoryear{Darling et al.}{2016}]{darling2016}Darling, J., Gerard, B., Amiri, N., Lawrence, K. 2016, \apj, 826, 24D
\bibitem[\protect\citeauthoryear{Felli et al.}{2007}]{felli2007}Felli, M., Brand, J., Cesaroni, R., et al. 2007, \aap, 476, 373F
\bibitem[\protect\citeauthoryear{Feretti et al.}{1998}]{feretti1998}Feretti, L., Giovannini, G., Klein, U., et al. 1998, \aap, 331, 475F
\bibitem[\protect\citeauthoryear{Fritz et al.}{2018}]{fritz2018}Fritz, T. K., Battaglia, G., Pawlowski, M. S., 2018, \aap, 619, 103F
\bibitem[\protect\citeauthoryear{Gillet et al.}{2015}]{gillet2015}Gillet, N., Ocvirk, P., \& Aubert, D., et al. 2015, \apj, 800, 34G
\bibitem[\protect\citeauthoryear{Gioia et al.}{1982}]{gioia1982} Gioia, I. M., Gregorini, L., Klein, U. 1982, \aap, 116, 164G
\bibitem[\protect\citeauthoryear{Govoni et al.}{2017}]{govoni2017}Govoni, F., Murgia, M., Vacca, V., et al. 2017, \aap, 603A, 122G
\bibitem[\protect\citeauthoryear{Greenhill et al.}{1990}]{greenhill1990}Greenhill, L. J., Moran, J. M., Reid, M. J., et al. 1990, \apj, 364, 513G 
\bibitem[\protect\citeauthoryear{Greenhill et al.}{1993}]{greenhill1993}Greenhill, L. J., Moran, J. M., Reid, M. J., Menten, K. M., Hirabayashi, H. 1993, \apj, 406, 482G
\bibitem[\protect\citeauthoryear{Helmi et al.}{2018}]{helmi2018}Gaia Collaboration, Helmi, A., van Leeuwen, F., McMillan, P. J., 2018, \aap, 616, 12G
\bibitem[\protect\citeauthoryear{Henkel et al.}{2018}]{henkel2018}Henkel, C., Greene, J.-E., \& Kamali, F. 2018, IAUS, 336, 69
\bibitem[\protect\citeauthoryear{Hindson et al.}{2018}]{hindson2018}Hindson, L., Kitchener, G., Brinks, E., et al. 2018, \apjs, 234, 29H 
\bibitem[\protect\citeauthoryear{Hopkins et al.}{2002}]{hopkins2002}Hopkins, A. M.,  Schulte-Ladbeck, R. E., \& Drozdovsky, I. O. 2002, \aj, 124, 862H
\bibitem[\protect\citeauthoryear{Ibata et al.}{2013}]{ibata2013}Ibata, R. A., Lewis, G. F., Conn, A. R., et al. 2013, \nat, 493, 62I
\bibitem[\protect\citeauthoryear{Klein \& Gr\"{a}ve}{1986}]{klein1986}Klein, U., \& Gr\"{a}ve, R. 1986, \aap, 161, 155K 
\bibitem[\protect\citeauthoryear{Karachentsev et al.}{2013}]{karachentsev2013}Karachentsev, I. D., Makarov, D. I., Kaisina, E. I., \apj, 145, 101
\bibitem[\protect\citeauthoryear{Lee et al.}{2005}]{lee2005}Lee, H., Skillman, E. D., Venn, K. A. 2005, \apj, 620, 223L
\bibitem[\protect\citeauthoryear{Loru et al.}{2019}]{loru2019}Loru, S., Pellizzoni, A., Egron, E., et al. 2019, \mnras, 482, 3857L
\bibitem[\protect\citeauthoryear{Mateo}{1998}]{mateo1998}Mateo, M. 1998, \araa, 36, 435M
\bibitem[\protect\citeauthoryear{McConnachie}{2012}]{mcconnachie2012}McConnachie, A. W. 2012, \aj, 144, 4M
\bibitem[\protect\citeauthoryear{Melis et al.}{2018}]{melis2018}Melis, A., Concu, R., Trois, A., et al. 2018, JAI, 750004M
\bibitem[\protect\citeauthoryear{Metz et al.}{2008}]{metz2008}Metz, M., Kroupa, P., \& Libeskind, N. I. 2008, \apj, 680, 287M
\bibitem[\protect\citeauthoryear{Millar \& Herbst}{1990}]{millar1990}Millar, T. J. \& Herbst, E.  1990, \mnras, 242, 92M
\bibitem[\protect\citeauthoryear{M\"{u}ller et al.}{2018}]{muller2018}M\"{u}ller, O., Pawlowski, M. S., Jerjen, H., et al. 2018, \sci, 359, 534
\bibitem[\protect\citeauthoryear{Murgia et al.}{2016}]{murgia2016}Murgia, M., Govoni, F., Carretti, E., et al. 2016, \mnras, 461, 3516M
\bibitem[\protect\citeauthoryear{Nishimura et al.}{2016}]{nishimura2016}Nishimura, Y., Shimonishi, T., Watanabe, Y., et al.  2016, \apj, 829, 94N
\bibitem[\protect\citeauthoryear{Panessa et al.}{2015}]{panessa2015}Panessa, F., Tarchi, A., Castangia, P., et al., 2015, \mnras, 447, 1289P  
\bibitem[\protect\citeauthoryear{Pawlowski et al.}{2012}]{pawlowski2012}Pawlowski, M. S., Pflamm-Altenburg, J., \& Kroupa, P. 2012, \mnras, 423, 1109P
\bibitem[\protect\citeauthoryear{Pawlowski et al.}{2013}]{pawlowski2013a}Pawlowski, M. S., Kroupa, P., \& Jerjen, H. 2013, \mnras, 435, 1928
\bibitem[\protect\citeauthoryear{Pawlowski \& Kroupa}{2013}]{pawlowski2013b}Pawlowski, M. S. \& Kroupa, P. 2013, \mnras, 435, 2116P
\bibitem[\protect\citeauthoryear{Pawlowski et al.}{2015}]{pawlowski2015}Pawlowski, M. S., McGaugh, S. S., \& Jerjen, H. 2015, \mnras, 453, 1047P
\bibitem[\protect\citeauthoryear{Perley et al.}{2013}]{perley2013}Perley, R. A., \& Butler, B. J. 2013, \apjs, 204, 19P
\bibitem[\protect\citeauthoryear{Pestalozzi et al.}{2007}]{pestalozzi2007}Pestalozzi, M. R., Chrysostomou, A., Collett, J. L., et al., 2007, \aap, 463, 1009P 
\bibitem[\protect\citeauthoryear{Prandoni et al.}{2017}]{prandoni2017}Prandoni, I., Murgia, M., Tarchi, A., et al. 2017, \aap, 608, 40
\bibitem[\protect\citeauthoryear{Quiroga-Nu\~nez et al.}{2017}]{quiroganunez2017}Quiroga-Nu\~nez, L. H., van Langevelde, H. J., Reid, M. J., Green, J. A. 2017, \aap, 604A, 72Q
\bibitem[\protect\citeauthoryear{Rubio et al.}{2015}]{rubio2015}Rubio, M., Elmegreen, B. G., Hunter, D. A., et al. 2015, \nat, 525, 218R
\bibitem[\protect\citeauthoryear{Schruba et al.}{2017}]{Schruba2017}Schruba, A., Leroy, A. K., Kruijssen, J. M. D., et al. 2017, \apj, 835, 278S
\bibitem[\protect\citeauthoryear{Sjouwerman et al.}{2010}]{sjouwerman2010}Sjouwerman, L. O., Murray, C. E., Pihlstr\"om, Y. M., Fish, V. L., Araya, E. D. 2010, \apj, 724L, 158S
\bibitem[\protect\citeauthoryear{Silich et al.}{2006}]{silich2006}Silich, S., Lozinskaya, T., Moiseev, A., et al. 2006, \aap, 448, 123S
\bibitem[\protect\citeauthoryear{Surcis et al.}{2011}]{surcis2011}Surcis, G., Vlemmings, W. H. T., Curiel, S., et al. 2011, \aap, 527, 48
\bibitem[\protect\citeauthoryear{Walterbos \& Braun}{1994}]{walterbos1994}Walterbos, R. A. M. \& Braun, R. 1994, \apj, 431, 156W
\bibitem[\protect\citeauthoryear{Watanabe \& Kouchi}{2002}]{watanabe2002}Watanabe, N. \& Kouchi, A. 2002, \apjl, 571, L173
\end{thebibliography}




\appendix

\section{Water maser tentative detections}\label{sect:tentative} 

In our search for maser sources, we considered as tentative detections those features that were detected at a signal-to-noise (SNR) level between a 4 and 7 $\sigma$ (see Sect.~\ref{sect:obs}). While no candidate detections were found at C band, a small number (12) of tentative water maser detections have been found, at K band, in each of the three galaxies, among which, one in IC\,1613, is seen at a 5.9-sigma level (Table \ref{table:allmasers}). The extremely low number of maser sources expected to be detectable in the galaxies (Sect.~\ref{sect:discussion}) makes, however, highly unlikely that all these detections are real. 


From a statistical point of view, we can estimate the probability that the 5.9-sigma tentative line detection, labeled IL4K, in IC\,1613, obtained by us in a single channel is spurious or not. Assuming a Gaussian distribution, the number of possibilities to have a n-sigma spurious event in a galaxy is given by  the number of positions/pixels (58$\times$58 or 28$\times$28, for NGC\,6822/IC\,1613 and WLM, respectively) times the number of channels (326) searched for a feature, times the probability to have a n-sigma spurious event. For a 5.9-sigma significance, the expected number of spurious detections are 0.004, 0.004, and 0.001, for NGC\,6822, IC\,1613, and WLM, respectively. For a 4 to 5 sigma significance, instead, the expected number of spurious detections ranges between 72 (4 sigma) and 0.65 (5 sigma), for NGC\,6822 or IC\,1613, and between 16 (4 sigma) and 0.15 (5 sigma) for WLM. Therefore, while some of the tentative detections at 4 to 5 sigma may indeed be expected to be factitious, the 5.9-sigma tentative detection can be considered as a significant maser candidate.

   \begin{figure}
   \includegraphics[width=8.5cm]{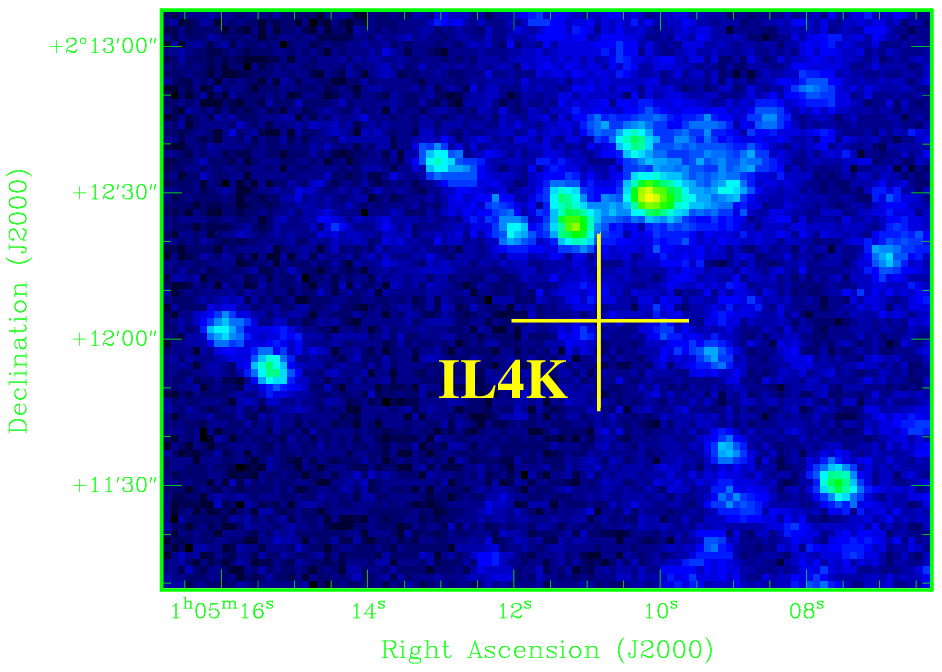}
      \caption{
Zoom-in of Fig.~\ref{fig:combo_ic1613}, showing the FUV Galex map of the region of IC\,1613 where the water maser 5.9-sigma tentative feature has been observed with the SRT. The yellow cross indicates the position of the feature (see Table~\ref{table:allmasers}).
}
         \label{fig:zoom_IL4K}
   \end{figure}

From an observational/astrophysical point of view, the IL4K water maser source line IC\,1613 is detected in a single channel of 92 kHz ($\sim$ 1.2 km/s at 22 GHz) with a peak flux density of 0.38 Jy, thus corresponding to an isotropic luminosity of about 0.005 \solum. The velocity of the feature is blueshifted w.r.t. the systemic velocity of the galaxy by $\sim$ 120 km/s (Table~\ref{table:allmasers}) and the location of the maser line does not coincide (although it is very close to) any obvious center of activity at FUV or radio wavelengths (see Fig.~\ref{fig:zoom_IL4K}). IL4K, because of its characteristics, may resemble IC\,10-NW, one of the two maser sources (somewhat unexpectedly; see Sect.~\ref{sect:discussion}) found in IC\,10 (\citealt{argon1994}; \citealt{baan1994}; \citealt{becker1993}), another LG dwarf galaxy. A main difference is, however, given by the large offset between the velocity of the maser line in IC\,1613 and the systemic velocity of the galaxy, that is not found in the IC10 source. Such a large offset cannot be justified by the gas velocity field of the galaxy derived by interferometric HI measurements that is relatively smooth and with a small velocity dispersion, with the exception of a few intermediate-size bubbles where an expansion velocity of 10-20 \kms was measured (\citealt{silich2006}). The possibility remains that mechanical energy released in the ISM from strong stellar winds, supernova explosions, or bubbles may influence the local velocity field through shocks.

Notably, two other weaker tentative features, labeled NL1K and NL5K, in NGC\,6822, have peculiarities that are worth to be mentioned. Indeed, NL1K is located in a particularly active star forming region to the NW, associated to NC1C (see Fig.~\ref{fig:combo_n6822}), and has a velocity offset w.r.t the systemic one of 'only' 52 \kms. The source NL5K has a much larger velocity offset ($\sim$ 120 \kms), but it is coincident with NC4aK, one of the two K band sources associated to NC4C, the strongest continuum source in the map (Fig.~\ref{fig:combo_n6822}).

Indeed, in order to try confirming the three aforementioned features, we re-observed at the positions of IL4K (on Jan. 15, 2019), and NL1K and NL5K (on Feb. 14, 2019) with the SRT (DDT project \#3-19)\footnote{It should be noted that the SRT was not available for observations from September 2016 to December 2018 because of extraordinary maintenance and subsequent technical/scientific recomissioning phases.} and with Effelsberg on Feb. 6 (only IL4K). In both cases, the observations were conducted in Position Switching mode with total on-source integration times of about 30 minutes (for the SRT) and 12 minutes (for Effelsberg), yielding a similar rms noise of $\sim$ 30 mJy/beam in a 25 and 38 kHz channel, for SRT and Effelsberg, respectively.
Oddly, and unfortunately, no line was detected at the velocity of the features originally detected in our ESP project above a 3-$\sigma$ level of 100 mJy/chan. Strong variability is indeed a possible explanation, given that narrow star-formation water maser features are typically highly variable (see, e.g., \citealt{felli2007}). However, the variability scenario would also imply a decrease of more than a factor 3-5 of the flux density in all three features within the same three-years period. Hence, the possibility that (some of) the tentative line emission features detected in our survey may be produced by backend artifacts or RFI, rather than having an astrophysical origin, cannot be confidently excluded. Additional follow-up observations of the other 9 tentative detections reported and/or a monitoring of the three aforementioned tentative targets would be therefore advisable. 

\begin{table*}
\caption{Tentative water maser features in NGC\,6822, IC\,1613, and WLM. The heliocentric velocity of the galaxies from \citet{mcconnachie2012}, V$_{Sys}$, are -57, -233, and -130 km/s, for  NGC\,6822, IC\,1613, and WLM, respectively. The uncertainty on the feature coordinates and velocities are given by the map cellsize (15$\arcsec$) and channel width (1.2 \kms), respectively. The uncertainty on the feature peak flux density is conservatively assumed to be the spectrum noise at the feature position (column 7).}             
\label{table:allmasers}      
\centering                          
\begin{tabular}{c c c c c c c c}        
\hline\hline     
Feature code    &      RA          &    Dec.  & V$_{Feat.}$ &  |V$_{Feat.}$-V$_{Sys}$| & Peak flux density & Spectral noise & SNR \\
          &      (J2000)     &  (J2000) &  km/s &   km/s  & (mJy)    &  (mJy)         &       \\ 
\hline 
\multicolumn{8}{c}{\bf {NGC\,6822}}\\
\hline
NL1K &    19 44 31.9 & -14 41 42  &      -5 &     52 &  543    &      127.4  &     4.3  \\ 
NL2K  &    19 44 37.0 & -14 55 12 &      15 &     72 &  384   &        85.8   &     4.5  \\  
NL3K &    19 44 40.1 & -14 48 12  &     -32 &     25 &  349    &       84.6  &      4.1  \\ 
NL4K &    19 44 46.3 & -14 46 12  &    -183 &    126 &  398    &       86.3  &      4.6  \\ 
NL5K  &   19 45 07.0 & -14 52 42 &      62 &    119 &   440    &      107.4  &     4.1  \\ 
\hline
\multicolumn{8}{c}{\bf {IC\,1613}}\\
\hline
IL1K  & 01:04:21.8  & 02:09:19 &   -99  &    134  &   276    &       66.4   &     4.2 \\
IL2K  & 01:04:22.8  & 02:13:49 &   -75  &    158  &   281    &       69.6   &     4.0 \\  
IL3K  & 01:04:52.8  & 02:04:04 &   -115 &    118  &   286    &       70.6   &     4.0 \\  
IL4K & 01:05:10.8  & 02:12:04 &   -352 &    119  &    384    &       65.0   &     5.9 \\  
\hline 
\multicolumn{8}{c}{\bf {WLM}}\\
\hline
WL1K   & 00 01 49.8  & -15 26 54  &    -152 &     22 &   663    &      162.0   &     4.1  \\ 		
WL2K  & 00 02 05.4  & -15 27 39  &    -139 &      9  &   660    &      159.7   &     4.1  \\ 		
WL3K  & 00 02 06.4  & -15 25 39  &    -299 &    169  &   691    &      162.1   &     4.3  \\		
\hline
\end{tabular}
\end{table*}
 

\bsp	
\label{lastpage}
\end{document}